\documentclass[12pt]{article} 
\usepackage[sectionbib]{natbib}
\usepackage{array,epsfig,rotating}
\usepackage[]{hyperref}  
\usepackage{sectsty, secdot}
\sectionfont{\fontsize{12}{14pt plus.8pt minus .6pt}\selectfont}
\renewcommand{\theequation}{\thesection\arabic{equation}}
\subsectionfont{\fontsize{12}{14pt plus.8pt minus .6pt}\selectfont}

\usepackage[left=2.5cm,right=2.5cm,top=2.5cm,bottom=3cm]{geometry}
\usepackage{algorithm}

\usepackage{amsmath}
\usepackage{amssymb}
\usepackage{amsfonts}
\usepackage{multirow}
\usepackage{amsthm}
\usepackage{comment}
\usepackage{xcolor}
\newtheorem{theorem}{Theorem}
\newtheorem{lemma}{Lemma}

\newtheorem{proposition}{Proposition}
\newtheorem{definition}{Definition}

\newtheorem{remark}{Remark}

\begin{document}

\global\long\def\cmdd{\textrm{CMD}_{\mathcal{H}}}%

\global\long\def\cmddn{\widehat{\textrm{CMD}}_{\mathcal{H}}}%

\global\long\def\mdd{\textrm{MD}_{\mathcal{H}}}%

\global\long\def\mddn{\widehat{\textrm{MD}_{\mathcal{H}}}}%

\global\long\def\mdcn{\widehat{\textrm{MC}_{\mathcal{H}}}}%

\global\long\def\myw{{Y}_{{\tau}}}%

\global\long\def\hatw{\hat{{Y}}_{{\tau}}}%

\global\long\def\mywk{y_{k_\tau}}%

\global\long\def\mdc{\textrm{MC}_{\mathcal{H}}}%

\global\long\def\cmdc{\textrm{CMC}_{\mathcal{H}}}%

\global\long\def\cmdcn{\widehat{\textrm{CMC}}_{\mathcal{H}}}%

\global\long\def\scmdc{\textrm{S-CMC}_{\mathcal{H}}}%


\renewcommand{\baselinestretch}{2}

\markright{ \hbox{\footnotesize\rm Statistica Sinica
}\hfill\\[-13pt]
\hbox{\footnotesize\rm
}\hfill }

\markboth{\hfill{\footnotesize\rm FIRSTNAME1 LASTNAME1 AND FIRSTNAME2 LASTNAME2} \hfill}
{\hfill {\footnotesize\rm FILL IN A SHORT RUNNING TITLE} \hfill}

\renewcommand{\thefootnote}{}
$\ $\par


\fontsize{12}{14pt plus.8pt minus .6pt}\selectfont \vspace{0.8pc}
\centerline{\large\bf Variable Screening via Conditional Martingale}
\vspace{2pt} 
\centerline{\large\bf Difference Divergence}
\vspace{.4cm} 
\centerline{Lei Fang$^{1}$, Qingcong Yuan$^{2}$, Xiangrong
Yin$^{1}$ and  Chenglong Ye$^{1}$}

\vspace{.4cm} 
\centerline{\it  $^1$Dr. Bing Zhang Department
of Statistics, University of Kentucky, $^2$Sanofi}
 \vspace{.55cm} \fontsize{9}{11.5pt plus.8pt minus.6pt}\selectfont


\begin{quotation}
\noindent {\it Abstract:}
Variable screening has been a useful research area that deals with ultrahigh-dimensional data. When there exist both marginally and jointly dependent predictors to the response, existing methods such as conditional screening or iterative screening often suffer from instability against the selection of the conditional set or the computational burden, respectively. In this article, we propose a new independence measure, named \textbf{c}onditional \textbf{m}artingale difference \textbf{d}ivergence ($\cmdd$), that can be treated as either a conditional or a marginal independence measure. Under regularity conditions, we show that the \textit{sure screening property} of $\cmdd$ holds for both marginally and jointly active variables. Based on this measure, we propose a kernel-based model-free variable screening method, which is efficient, flexible, and stable against high correlation among predictors and heterogeneity of the response. In addition, we provide a data-driven method to select the conditional set. In simulations and real data applications, we demonstrate the superior performance of the proposed method.
\vspace{9pt}

\par
\end{quotation}\par

\def\thefigure{\arabic{figure}}
\def\thetable{\arabic{table}}

\renewcommand{\theequation}{\thesection.\arabic{equation}}

\fontsize{12}{14pt plus.8pt minus .6pt}\selectfont

\section{ Introduction\label{sec:intro}}

\sloppy Variable screening has been a research area that deals with ultrahigh-dimensional data, where high-dimensional methods may fail due to the curse of dimensionality, as \citet{fan2009ultrahigh} suggested.
\citet{fan2008sure}'s seminal work suggests to screen the ultrahigh-dimensional
data before conducting variable selection. They proposed a sure independent
screening (SIS) method for linear models to screen out inactive
variables based on Pearson correlation. After that, variable screening receives more attention since it only requires that the selected set
of variables covers the set of active variables, which is referred
to as the \textit{sure screening property} \citep{fan2008sure}. Screening methods with this property
suffer less from  \textit{instability} \citep{10.3150/13-BEJSP14} that is seen in many variable selection
methods.

Various model-based screening methods
have been developed. For linear regression models, screening methods have been
proposed based on different measures, including marginal Pearson correlation
\citep{fan2008sure}, forward regression \citep{wang2009forward},
marginal empirical likelihood ratio \citep{10.1214/13-AOS1139}, and
Kendall's rank correlation \citep{10.1214/12-AOS1024}. For linear
quantile regression, screening methods based on quantile partial correlation
\citep{doi:10.1080/01621459.2016.1156545} and conditional quantile
correlation \citep{ZHANG20181} have been proposed to handle heterogeneous
data. In the context of generalized linear models, screening methods
based on the maximum marginal likelihood or its estimate \citep{10.1214/10-AOS798},
the sparsity-restricted maximum likelihood estimator \citep{doi:10.1080/01621459.2013.879531},
and Kolmogorov-Smirnov statistic \citep{10.2307/43304549} have also
been proposed. Other screening methods include model settings such as linear
regression models with interactions \citep{hao2014interaction,fan2016interaction,kong2017interaction},
Cox models \citep{ZHAO2012397}, varying coefficient models \citep{10.1214/14-AOS1236,10.2307/24310967},
and additive models \citep{doi:10.1198/jasa.2011.tm09779}.

An alternative approach is the model-free screening method, which recently
has gained popularity due to its less stringent assumptions. \citet{doi:10.1198/jasa.2011.tm10563}
proposed a sure independence ranking and screening approach (SIRS)
for index models. \citet{li2012feature} proposed a screening method
(DC-SIS) based on distance correlation, which can be applied to
grouped variables. \citet{doi:10.1080/01621459.2014.920256} proposed
a model-free screening method (MV-SIS) based on empirical conditional distribution
function for discriminant analysis. \citet{shao2014martingale} proposed
the use of martingale difference correlation (MDC), which can be applied
to mean and quantile screening. \citet{10.1214/14-AOS1303} proposed
the fused Kolmogorov filter that works with different types of response
variables and high covariate correlation. \citet{doi:10.1080/01621459.2020.1864380}
proposed a screening method based on covariate information number (CIN)
motivated by Fisher information. \citet{10.1214/18-AOS1738}
developed a screening framework from the perspective of loss functions
and proposed a screening method based on conditional strictly convex
losses. Based on \textit{ball correlation}, \citet{doi:10.1080/01621459.2018.1462709} proposed a generic
screening method for biomedical discovery.

As pointed out by \citet{li2012feature,shao2014martingale,tian2021rase}, screening methods based on marginal measures (e.g., the marginal correlation between the response and each predictor) will possibly
miss the marginally but not jointly independent predictors. Two types of approaches have been developed to handle this issue. One approach
is {\it  conditional screening}, a screening procedure based on a given set of variables. For example,  \citet{barut2016conditional} proposed conditional
sure independence screening (CSIS) for assessing the conditional (on a given conditional set) contribution of a predictor to the response in generalized linear models. Based on conditional distance correlation, \citet{wen2018sure} proposed
a method that adjusts for confounding variables. \cite{tong2022model} proposed a conditional independence measure and its corresponding screening method (CIS) with false discovery rate (FDR) control, which also works for heavy-tailed predictors/responses. Another approach
is {\it  screening via iterative procedures.} For example, the aforementioned forward
regression iteratively selects the variables. \citet{doi:10.1080/01621459.2013.879531}
considered a method based on sparse MLE, where the algorithm iteratively
updates the coefficients in the link function. \citet{zhou2020model}
proposed a model-free forward screening method that iteratively updates
the conditional set and is robust to outliers. \citet{tian2021rase}
proposed an iterative variable screening method based on random subspace
ensembles (RaSE) with a theoretical guarantee for iterative
screening procedures. 

However, two challenges remain. Model-based iterative methods (e.g., iterative SIS) may
rely on a specific variable selection method, which makes the procedure less stable. On the other hand,
the conditional screening method requires prior knowledge of the conditional set and its performance becomes unstable if an unreasonable conditional set is selected. It motivates
us to develop a stable model-free screening method that identifies both marginally and jointly dependent variables to the response. We propose a kernel-based measure that captures both conditional
and marginal mean-independent relationships. In
particular, via Bochner's theorem \citep{wendland2004scattered},
we transform the problem of choosing weights, a key element in our independence measure, to the problem of choosing
kernels and their bandwidths in reproducing kernel Hilbert space (RKHS). This flexible kernel-based fashion allows our method to perform well in various settings, as illustrated in the synthetic and real data analysis.

The advantages of our method are as follows. First, we propose a kernel-based
independence measure ($\cmdd$) that is able to characterize both conditional and marginal mean independence. Thus,
we propose a $\cmdd$-based screening method that can detect both marginally
and jointly dependent/active variables. Second, the proposed model-free
screening method is stable against outliers, data heterogeneity, and
high covariate correlation. Third, we show the sure screening property holds for screening both marginally and jointly dependent variables under mild regularity conditions.
We also suggest selecting a data-driven conditional set for conducting conditional screening when no prior information
is available.

The rest of the article is organized as follows. Section \ref{section:background}
introduces the proposed independence measure and its theoretical properties.
In Section \ref{section:definition:MDCscreening}, we propose a model-free
variable screening procedure and present its sure screening property.
The simulation results and two real data examples are reported in
Section \ref{section: simulation}, followed by the conclusion in
Section \ref{section:conclusion}. Additional theorems are presented in the appendix.  Auxiliary simulation results and technical proofs are included in the supplementary material.
\section{General methodology\label{section:background}}

\paragraph{Notations.}

Throughout the article, we use upper case (e.g., $V$) to denote a random variable and use
bold font to denote a random vector (e.g., $\boldsymbol{U}$).  We use $(\boldsymbol{U}_1',\boldsymbol{U}_2',V')$ and $(\boldsymbol{U}_1'',\boldsymbol{U}_2'',V'')$ to denote $i.i.d$
copies of $(\boldsymbol{U}_1,\boldsymbol{U}_2,V)$. For a complex function $f(\boldsymbol{s}):\mathbb{R}^q\rightarrow \mathbb{C}^p$, denote its RKHS norm as $||f(\boldsymbol{s})||^2_{\mathcal{H}_k}=\int_{\mathbb{R}^q}|f(\boldsymbol{s})|^2w(\boldsymbol{t})d\boldsymbol{t}$, where $w(\boldsymbol{t})$ corresponds to the kernel function $k$ in RKHS.

For a sequence $\{t_{ij}\}$ with double
indices $i,j=1,...,n$, we define 
\begin{equation}
t_{ij}^{*}=t_{ij}-\bar{t}_{i.}-\bar{t}_{.j}+\bar{t}_{..},\label{eq:star-1}
\end{equation}
where $\bar{t}_{.j}=\frac{1}{(n-2)}\sum_{i=1}^{n}t_{ij}$, $\bar{t}_{i.}=\frac{1}{(n-2)}\sum_{j=1}^{n}t_{ij}$,
and $\bar{t}_{..}=\frac{1}{(n-1)(n-2)}\sum_{i=1}^{n}\sum_{j=1}^{n}t_{ij}$.
Denote $\langle\boldsymbol{a},\boldsymbol{b}\rangle$ as the inner
product of any two vectors $\boldsymbol{a},\boldsymbol{b}$ of the
same dimension. 

\subsection{Conditional Martingale Difference
Divergence}

\paragraph{A motivating example.}

Assume two random variables $X_{1}$ and $X_{2}$ are independent
with $E(X_{2})=0$, and let $Y=X_{1}X_{2}$. We have $E(Y|X_{1})-E(Y)=E(X_{2})\cdot(X_{1}-E(X_{1}))=0$
as long as $E(X_{2})=0$. The condition ``$E(X_{2})=0$" is mild as we can standardize predictors in the dataset to have mean 0
in practice. This example indicates that mean independence measures
based on the relationship $E(Y|X_{1})-E(Y)$ will misleadingly suggest
that $X_{1}$ is independent from mean of $Y$. It was pointed out by \citet{li2012feature}
and \citet{shao2014martingale} that the marginal measures/methods such as DC and MDC will possibly miss the variables that
only jointly contribute to the response variable. It motivates us
to consider the following: 
\[
E(Y|X_{1},X_{2})-E(Y|X_{2}),
\]
which equals to $X_{2}\cdot(X_{1}-E(X_{1}))$ in this example. More generally, we consider
the equality $E(V|\boldsymbol{U}_{1},\boldsymbol{U}_{2})=E(V|\boldsymbol{U}_{1})$, i.e., the response variable $V$ and covariate vector $\boldsymbol{U}_{2}$
given the covariate vector $\boldsymbol{U}_{1}$. If in addition, $\boldsymbol{U}_{1}$
and $\boldsymbol{U}_{2}$ are independent, we have $E(V|\boldsymbol{U}_{1},\boldsymbol{U}_{2})=E(V|\boldsymbol{U}_{1})$
if and only if $E(Ve^{i\langle \boldsymbol{t}_{1},\boldsymbol{U}_{1}\rangle}|\boldsymbol{U}_{2})=E(Ve^{i\langle \boldsymbol{t}_{1},\boldsymbol{U}_{1}\rangle})$ for any $\boldsymbol{t}_1$ (the proof is presented in supplementary material S2 (b)).
This motivates us to propose the following independence measure in Definition \ref{def: CMDD}, which can be treated as either a conditional
or a marginal (see Remark \ref{rmk2}) mean independence measure.
\begin{definition}\label{def: CMDD} Given a random vector $\boldsymbol{U}_{1}  \in \mathbb{R}^p$,
the \textbf{c}onditional \textbf{m}artingale difference \textbf{d}ivergence
of a random variable $V$ and a random vector $\boldsymbol{U}_{2} \in \mathbb{R}^q$
is defined as 
\begin{align*}
\label{eq:CMC-integral}
&\textnormal{\ensuremath{\cmdd}}^{2}(V,\boldsymbol{U}_{2}|\boldsymbol{U}_{1})
\\&=\iint|E(Ve^{i(\langle \boldsymbol{t}_{1},\boldsymbol{U}_{1}\rangle+\langle \boldsymbol{t}_{2},\boldsymbol{U}_{2}\rangle)})-E(Ve^{i\langle \boldsymbol{t}_{1},\boldsymbol{U}_{1}\rangle})E(e^{i\langle \boldsymbol{t}_{2},\boldsymbol{U}_{2}\rangle})|^{2}w_1(\boldsymbol{t}_{1})w_2(\boldsymbol{t}_{2})d\boldsymbol{t}_{1}d\boldsymbol{t}_{2},
\end{align*}
where $w_1(\boldsymbol{t}_{1})$ and $w_2(\boldsymbol{t}_{2})$ are  weight functions.
\end{definition}

The measure $\cmdd$ depends on two ingredients:
a mean independence measure of a random vector and a random variable, and an adjusting method
of the effect of a third vector. It provides valuable information
on the conditional contribution of $\boldsymbol{U}_2$ to the mean of $V$ given $\boldsymbol{U}_1$.
\begin{remark} \label{rmk:kernel}We choose the weight function $w_1(\boldsymbol{t}_{1})$ and $w_2(\boldsymbol{t}_{2})$
to be integrable, relaxing the strong assumption of the boundedness
of $\boldsymbol{U}_{1}$ and $\boldsymbol{U}_{2}$ as in the literature. The choice of an
integrable weight function makes the proposed independence
measure more flexible. In particular,
we can rewrite $\textnormal{\ensuremath{\cmdd}}(V,\boldsymbol{U}_{2}|\boldsymbol{U}_{1})$ as a functional of kernel functions in RKHS. See more details in Theorem \ref{theorem:CMDC}(a) and Remark \ref{remark:weight}.
\end{remark} 
We now define a scale-invariant version of the proposed measure.
\begin{definition} \sloppy Let $k_1$ and $k_2$ be the two kernel functions that correspond to the weight function $w_1(\boldsymbol{t}_1)$ and $w_2(\boldsymbol{t}_2)$, as illustrated in Theorem \ref{theorem:CMDC}(a). We define the \textbf{c}onditional \textbf{m}artingale difference \textbf{c}orrelation
\begin{equation*}
\textnormal{\ensuremath{\cmdc}}(V,\boldsymbol{U}_{2}|\boldsymbol{U}_{1})=\begin{cases}
\dfrac{\textnormal{\ensuremath{\cmdd}}(V,\boldsymbol{U}_{2}|\boldsymbol{U}_{1})}{\sqrt{\textrm{v}(k_{2},\boldsymbol{U}_{2}){v}(k_{1_V},\boldsymbol{U}_{1})}} & \text{if  $\textrm{v}(k_{2},\boldsymbol{U}_{2}){v}(k_{1_V},\boldsymbol{U}_{1})>0$ }\\
0 & \text{ otherwise,}
\end{cases}
\end{equation*}
where \sloppy $\textrm{v}(k,\boldsymbol{U}):=E{[}k^{2}(\boldsymbol{U},\boldsymbol{U}'){]}+E^{2}{[}k(\boldsymbol{U},\boldsymbol{U}'){]}-2E[k(\boldsymbol{U},\boldsymbol{U}^{'})\cdot k(\boldsymbol{U},\boldsymbol{U}^{''})]$
and $k_{V}(\boldsymbol{U},\boldsymbol{U}'):=VV^{'}k(\boldsymbol{U},\boldsymbol{U}')$ for any kernel function $k$.\end{definition}

\begin{remark} \label{rmk2}When $\boldsymbol{U}_{1}$ contains no useful information
($\boldsymbol{U}_{1}=\emptyset$, $\boldsymbol{U}_{1}\equiv\mathbf{c}$,
or $\boldsymbol{U}_{1}$ is independent from $(V,\boldsymbol{U}_{2})$),
the definition of $\textnormal{\ensuremath{\cmdd}}$ reduces to a marginal mean independence measure.
That is,
\begin{align*}
\textnormal{\ensuremath{\mdd}}^{2}(V,\boldsymbol{U}_{2}):=\int|E(Ve^{i\langle \boldsymbol{t}_{2},\boldsymbol{U}_{2}\rangle})-E(V)E(e^{i\langle \boldsymbol{t}_{2},\boldsymbol{U}_{2}\rangle})|^{2}w(\boldsymbol{t}_{2})d\boldsymbol{t}_{2}.
\end{align*}
Note that $\textnormal{\ensuremath{\mdd}}(V,\boldsymbol{U}_{2})$ is a generalized version of \textnormal{MDD} \citep{shao2014martingale}
by kernerlizing the $L_2$ distance of $\boldsymbol{U}_2$ and its i.i.d. copy $\boldsymbol{U}'_2$. {The standardized version \textnormal{\ensuremath{\mdc}} is defined similarly as  \textnormal{\ensuremath{\cmdc}}}.
\end{remark} 

As will be seen in Theorem \ref{theorem:CMDC}, the
definition of the $\cmdc$ is more convenient for variable screening
purpose since it takes values in $[0,1]$. More detailed {discussion} of $\cmdc$
is included in Section \ref{section:definition:MDCscreening}. Now we show the theoretical properties of the proposed conditional independence
measure. 

\begin{theorem}\label{theorem:CMDC} Assume $E(V^{2})<\infty$, we have the following properties:%

\begin{enumerate}
\item[(a).] 
We can rewrite $\textnormal{\ensuremath{\cmdd}}^2{(V,\boldsymbol{U}_{2}|\boldsymbol{U}_{1})}$ as
\begin{align*}
 & \textnormal{\ensuremath{\cmdd}}^{2}(V,\boldsymbol{U}_{2}|\boldsymbol{U}_{1})\\
 & =E(VV'k_{1}(\boldsymbol{U}_{1},\boldsymbol{U}_{1}^{'})k_{2}(\boldsymbol{U}_{2},\boldsymbol{U}_{2}^{'}))+E(VV'k_{1}(\boldsymbol{U}_{1},\boldsymbol{U}_{1}^{'}))E(k_{2}(\boldsymbol{U}_{2},\boldsymbol{U}_{2}^{'}))\\
 & -2E(VV'k_{1}(\boldsymbol{U}_{1},\boldsymbol{U}_{1}^{'})k_{2}(\boldsymbol{U}_{2},\boldsymbol{U}_{2}^{''})),
\end{align*}
where $k_{1}$ and $k_{2}$ are RHKS
kernel functions determined by $w_1(\boldsymbol{t}_{1})$ and $w_2(\boldsymbol{t}_{2})$ defined in Definition \ref{def: CMDD}, respectively.
\item[(b).] $0\le\textnormal{\ensuremath{\cmdc}}(V,\boldsymbol{U}_{2}|\boldsymbol{U}_{1})\le1,$
and $\textnormal{\ensuremath{\cmdc}}(V,\boldsymbol{U}_{2}|\boldsymbol{U}_{1})=0\Leftrightarrow E(V|\boldsymbol{U}_{1},\boldsymbol{U}_{2})=E(V|\boldsymbol{U}_{1})$
a.s. if $\boldsymbol{U}_{1} \perp \boldsymbol{U}_{2}$. 
\item[(c).] Given two constants $d\in\mathbb{R}$ and $e\in\mathbb{R}$, $\textnormal{\ensuremath{\cmdc}}(a+bV,\boldsymbol{c}+d\boldsymbol{U}_{2}|e\boldsymbol{U}_{1})=\textnormal{\ensuremath{\cmdc}}(V,\boldsymbol{U}_{2}|\boldsymbol{U}_{1})$
for any scalars $a,b\in\mathbb{R}$ and $\boldsymbol{c}\in\mathbb{R}^{q}$. {If the kernels $k_1$ and $k_2$ in (a) are scale-invariant, the above equality holds for any scalars $d$ and $e$ as well}. 

\item[(d).] If the random variables $U,V\in\mathbb{R}$ are independent, then
\[
\textnormal{\ensuremath{\mdc}}^{2}(VU,U)=\frac{E^{2}(V)}{Var(V)+E^{2}(V)+E^{2}(U)\frac{Var(V)}{Var({U})}}\textnormal{\ensuremath{\mdc}}^{2}(U,U).
\]

Furthermore, if $E(V)=0$, then $\textnormal{\ensuremath{\mdc}}(VU,U)=0$. 
\end{enumerate}
\end{theorem}
\begin{remark}\label {remark:weight}
{If we take non-integrable weight functions $w_1(\boldsymbol{t}_{1})$ and $w_2(\boldsymbol{t}_{2})$ in Definition \ref{def: CMDD}, then $k_1$ and $k_2$ in Theorem 1(a) may not be translation-invariant kernels in RKHS (e.g., the Euclidean distance function). See dCov \citep{szekely2007measuring} for an example that adopts a non-integrable weight in its definition.}

\end{remark}

\begin{remark}\label{rmk5}
Property (b)
shows the equivalence between the conditional mean independence and $\textnormal{\ensuremath{\cmdc}}$
being 0, which suggests $\textnormal{\ensuremath{\cmdc}}$ is a suitable tool for conducting
variable screening. Note that the independence of $\boldsymbol{U}_1$ and $\boldsymbol{U}_2$ in Property (b) is to ease the proof. Indeed, if we define a new independence measure $\textnormal{\ensuremath{\cmdd}}_{,new}^{2}(V,\boldsymbol{U}_{2}|\boldsymbol{U}_{1})=\iint|E(Ve^{i(\langle \boldsymbol{t}_{1},\boldsymbol{U}_{1}\rangle+\langle \boldsymbol{t}_{2},\boldsymbol{U}_{2}\rangle)}|\boldsymbol{U}_{1})-E(Ve^{i\langle \boldsymbol{t}_{1},\boldsymbol{U}_{1}\rangle}|\boldsymbol{U}_{1})E(e^{i\langle \boldsymbol{t}_{2},\boldsymbol{U}_{2}\rangle}|\boldsymbol{U}_{1})|^{2}w_1(\boldsymbol{t}_{1})w_2(\boldsymbol{t}_{2})d\boldsymbol{t}_{1}d\boldsymbol{t}_{2},$
then $$\textnormal{\ensuremath{\cmdd}}_{,new}(V,\boldsymbol{U}_{2}|\boldsymbol{U}_{1})=0 \textit{ a.s.}\Leftrightarrow E(V|\boldsymbol{U}_{1},\boldsymbol{U}_{2})=E(V|\boldsymbol{U}_{1})\textit{ a.s..}$$ 
This removes the independence condition of $\boldsymbol{U}_1$ and $\boldsymbol{U}_2$. Then we need to replace our $U$-statistics estimator with the conditional $U$-statistics to estimate the new independence measure. Note that this new measure $\textnormal{\ensuremath{\cmdd}}_{,new}$ is a function of the random vector $\boldsymbol{U}_{1}$. Such a conditional measure and its associated screening method is left as future work of interest. In this article, we stick to our original proposed measure $\textnormal{\ensuremath{\cmdd}}$. In the simulations, as we see, even $\boldsymbol{U}_1$ and $\boldsymbol{U}_2$ are not independent (e.g., high variable correlations $\rho=0.5, 0.8,0.9$ in Example 1, and nonlinearly associated predictors in Example 4), our variable screening method still performs well, or even outperforms other methods in almost all the simulation settings.
\end{remark}

\begin{remark} Property (c) shows that the proposed $\cmdc$ is scale-invariant. Property (d) directly shows the deficiency of marginal-type mean independence 
measure $(\textnormal{\ensuremath{\mdc}})$ in interaction screening. Thus we propose the variable screening procedure based on $\textnormal{\ensuremath{\cmdc}}$. \end{remark}

\subsection{Empirical Estimators and Asymptotic Properties}

Based on property (a) in Theorem \ref{theorem:CMDC}, we construct the
$U$-statistic to estimate $\cmdc$. 

\begin{definition} Let $(\boldsymbol{U}_{1i},\boldsymbol{U}_{2i},V_{i})_{i=1}^{n}$
be $i.i.d.$ observations of $(\boldsymbol{U}_{1},\boldsymbol{U}_{2},V)$.
Denote $a_{ij}=V_{i}V_{j}k_{1}(\boldsymbol{U}_{1i},\boldsymbol{U}_{1j})$
and $b_{ij}=k_{2}(\boldsymbol{U}_{2i},\boldsymbol{U}_{2j})$ for $i,j=1,...,n$.
Define the corresponding $a_{ij}^{*}$ and $b_{ij}^{*}$ as in Equation
\eqref{eq:star-1}. The $U$-statistic estimator of $\textnormal{\ensuremath{\cmdd}}$ is
\[
\textnormal{\ensuremath{\cmddn}}(V,\boldsymbol{U}_{2}|\boldsymbol{U}_{1})=\frac{1}{n(n-3)}\sum_{1\le i\ne j\le n}a_{ij}^{*}b_{ij}^{*},
\]
 and the corresponding estimator of
 $\textnormal{\ensuremath{\cmdc}}$ is: 
\[
\textnormal{\ensuremath{\cmdcn}}(V,\boldsymbol{U}_{2}|\boldsymbol{U}_{1})=\dfrac{\sum_{1\le i\ne j\le n}a_{ij}^{*}b_{ij}^{*}}{\sqrt{\sum_{1\le i\ne j\le n}a_{ij}^{*2}\sum_{1\le i\ne j\le n}b_{ij}^{*2}}}.
\]
\end{definition}

\begin{remark} Compared to the adoption of $V$-statistic estimator, we choose the $U$-statistic
because it is unbiased and less computationally expensive.
\end{remark}

We now show the strong consistency of the proposed estimators.

\begin{theorem} (Consistency) If E $(V^{2})<\infty$, then
\begin{equation}
lim_{n\rightarrow\infty}\textnormal{\ensuremath{\cmddn}}(V,\boldsymbol{U}_{2}|\boldsymbol{U}_{1})=\textnormal{\ensuremath{\cmdd}}(V,\boldsymbol{U}_{2}|\boldsymbol{U}_{1})\textrm{ a.s.,}
\end{equation}
and 
\begin{equation}
lim_{n\rightarrow\infty}\textnormal{\ensuremath{\cmdcn}}(V,\boldsymbol{U}_{2}|\boldsymbol{U}_{1})=\textnormal{\ensuremath{\cmdc}}(V,\boldsymbol{U}_{2}|\boldsymbol{U}_{1})\textrm{ a.s.}.
\end{equation}
\end{theorem}

In the next theorem, we derive the asymptotic distribution for $\textnormal{\ensuremath{\cmdd}}(V,\boldsymbol{U}_{2}|\boldsymbol{U}_{1})$.
Denote the following functions: $\textsl{g}_{\boldsymbol{U}_2}(\boldsymbol{t}_2):=E(e^{i\langle\boldsymbol{t}_2,\boldsymbol{U}_2\rangle})$, $\textsl{g}_{V,\boldsymbol{U}_1}(\boldsymbol{t}_1):=E(Ve^{i\langle\boldsymbol{t}_1,\boldsymbol{\boldsymbol{U}_1}\rangle})$, and $F(\boldsymbol{t}_{1},\boldsymbol{t}_{2}):=E(V^{2}e^{i\langle\boldsymbol{t}_{1},\boldsymbol{U}_{1}\rangle}e^{i\langle\boldsymbol{t}_{2},\boldsymbol{U}_{2}\rangle})$. Define the covariance function $\textnormal{cov}_{\Gamma}((\boldsymbol{t}{}_{1},\boldsymbol{t}{}_{2}),(\boldsymbol{t}'_{1},\boldsymbol{t}'_{2})):=F(\boldsymbol{t}_{1}-\boldsymbol{t}'_{1},\boldsymbol{t}_{2}-\boldsymbol{t}'_{2})+(F(\boldsymbol{t}_{1}-\boldsymbol{t}'_{1},0)+\textsl{g}_{V,\boldsymbol{U}_{1}}(\boldsymbol{t}{}_{1})\overline{\textsl{g}_{V,\boldsymbol{U}_{1}}(\boldsymbol{t}'_{1})})\{\textsl{g}_{\boldsymbol{U}_{2}}(\boldsymbol{t}{}_{2})\overline{\textsl{g}_{\boldsymbol{U}_{2}}(\boldsymbol{t}'_{2})}-\textsl{g}_{\boldsymbol{U}_{2}}(\boldsymbol{t}_{2}-\boldsymbol{t}'_{2})\}$$-F(\boldsymbol{t}_{1}-\boldsymbol{t}'_{1},\boldsymbol{t}_{2})\overline{\textsl{g}_{\boldsymbol{U}_{2}}(\boldsymbol{t}'_{2})}-F(\boldsymbol{t}_{1}-\boldsymbol{t}'_{1},-\boldsymbol{t}'_{2})\textsl{g}_{\boldsymbol{U}_{2}}(\boldsymbol{t}{}_{2})$.\begin{theorem}
 Assume  $E(V^{2})<\infty$, we have the following:
\begin{enumerate}
\item[a.] If $\textnormal{\ensuremath{\cmdd}}(V,\boldsymbol{U}_{2}|\boldsymbol{U}_{1})=\textnormal{0}$,
then 
\begin{equation}
n\textnormal{\ensuremath{\cmddn}}^{2}(V,\boldsymbol{U}_{2}|\boldsymbol{U}_{1})\overset{d}{\rightarrow}||\Gamma(s)||_{\mathcal{H}_k}^{2}
\end{equation}
as $n\rightarrow\infty$, where $\Gamma(\cdot)$ is a complex-valued
zero-mean Gaussian random process with covariance function $\textnormal{cov}_{\Gamma}((\boldsymbol{t}{}_{1},\boldsymbol{t}{}_{2}),(\boldsymbol{t}'_{1},\boldsymbol{t}'_{2}))$.
\item[b.] If $\textnormal{CMD\ensuremath{_{\mathbf{\mathcal{H}}}}}(V,\boldsymbol{U}_{2}|\boldsymbol{U}_{1})=0$
and $E(V^{2}|\boldsymbol{U}_{2})=E(V^{2})$, then 
\[
n\textnormal{\ensuremath{\cmddn}}^{2}(V,\boldsymbol{U}_{2}|\boldsymbol{U}_{1})/S_{n}\overset{d}{\rightarrow}\sum_{j=1}^{\infty}\lambda_{j}Z_{j}
\]
as $n\rightarrow\infty$, where $S_{n}=(\frac{1}{n}\sum_{i}V_{i}^{2}-\frac{1}{n(n-1)}\sum_{i\ne j}a_{ij})(1-\frac{1}{n(n-1)}\sum_{i\ne j}b_{ij})$,
$Z_{j}\overset{i.i.d.}{\sim}\chi_{1}^{2}$, and $\{\lambda_{j}\}_{j=1}^{\infty}$
are nonnegative constants such that $E(\sum_{j=1}^{\infty}\lambda_{j}Z_{j})=1$. 
\item[c.] If $\textnormal{\ensuremath{\cmdd}}(V,\boldsymbol{U}_{2}|\boldsymbol{U}_{1})>0$,
then $n\cdot\textnormal{\ensuremath{\cmddn}}^{2}(V,\boldsymbol{U}_{2}|\boldsymbol{U}_{1})/S_{n}\overset{p}{\rightarrow}\infty$
as $n\rightarrow\infty$. 
\end{enumerate}
\end{theorem}

The properties stated in the theorems of this section motivate us to propose variable screening algorithm based on $\cmdc$ and its estimate $\cmdcn$.

\section{$\protect\cmdc$-based Variable Screening\label{section:definition:MDCscreening}}

In this section, we show the sure screening property of $\cmdc$ in Section
\ref{subsec:Sure-screening-property}. In Section \ref{subsec:S-screen},
we introduce a variable screening algorithm named $\scmdc$ to accommodate the dependence among predictors. The algortihm works reasonably well when the data suffer from outliers, high correlation, and heterogeneity as seen in the numerical studies. 
\subsection{Sure Screening Property\label{subsec:Sure-screening-property}}
Without loss of generality, let $Y$ be a univariate continuous response
variable and $\boldsymbol{X}=(X_{1},...,X_{p})^{T}$ be the predictor
vector. Denote the sample as $(X_{1k},...,X_{pk},Y_{k})_{k=1}^{n}$,
where $p\gg n$. For any index set $S\subseteq\{1,...,p\}$,
denote $\boldsymbol{X}_{S}:=\{\boldsymbol{X}_{j}:j\in S\}$. Given a conditional set $\boldsymbol{X}_{S}$ with cardinality $d_1$, we define 
\[
\mathcal{D}_{S}=\{j:\mathbb{E}(Y|(\boldsymbol{X}_{S},X_j))\ \text{depends on}\ X_{j}\}
\]
as the index set of dependent/active predictors conditional on $\boldsymbol{X}_{S}$, and 
\[
\mathcal{I}_{S}=\{j:\mathbb{E}(Y|(\boldsymbol{X}_{S},X_j))\ \text{is independent from}\ X_{j}\}
\]
as the index set of independent/inactive predictors conditional on $\boldsymbol{X}_{S}$. Note that $\mathcal{D}_{S}$ is a subset of $\mathcal{D}:=\{j:\mathbb{E}(Y|\boldsymbol{X})\ \text{depends on}\ X_{j}\}$, the set of all dependent predictors. Suppressing $S$, we
denote $\omega_{j}=\cmdc^2(Y,X_{j}|\boldsymbol{X}_{S})$ as the
dependence score of $X_{j}$ given $\boldsymbol{X}_{S}$.
Let $\hat{\omega}_{j}=\widehat{\cmdc}^2(Y,X_{j}|\boldsymbol{X}_{S})$
be the estimator of $\omega_{j}$ and 
\[
\hat{\mathcal{D}}_{S}=\{j:\hat{\omega}_{j}\ge cn^{-\kappa},\ \text{for}\  j\in S^c\}
\]
be the set of selected variables after screening. Before stating the sure screening property, we assume the following conditions. 

\begin{description}
\item [{(A1)}] There exists a constant $s_{0}>0$ such that $E(\exp(sY^{2}))<\infty$
for all $0<s\le2s_{0}$.
\item [{(A2)}] For any given $\boldsymbol{X}_{S}$, $\min_{j\in\mathcal{D}_{S}}\omega_{j}\ge2cn^{-\kappa}$
for some constant $c>0$ and $0\le\kappa<1/2$. 
\end{description}
\sloppy Condition $(A1)$ puts constraint on the tail distribution of the response variable and Condition (A2) requires that the conditional active/dependent variables and inactive/independent variables are well separated.
\begin{theorem} \label{Thm:sure}Under Condition $(A1)$, for any $0<\gamma<1/2-\kappa$,
there exist positive constants $c_{1}$ and $c_{2}$ such that
\begin{equation}\label{eq:7}
P\Bigg\{\underset{1\leq j\leq p-d_1}{\max}|\widehat{\omega}_{j}-\omega_{j}|\ge cn^{-\kappa}\Bigg\}\le O((p-d_1)[\exp(-c_{1}n^{1-2(\kappa+\gamma)})+n\exp(-c_{2}n^{\gamma})]).
\end{equation}
If Conditions $(A2)$ also holds, we have
\begin{equation}
P(\mathcal{D}_{S}\subset\hat{\mathcal{D}}_{S})\ge1-O(s_{n}[\exp(-c_{1}n^{1-2(\kappa+\gamma)})+n\exp(-c_{2}n^{\gamma})])
\label{eq8}\end{equation}
for any conditional set $\boldsymbol{X}_S$, where $s_{n}$ is the cardinality of $\mathcal{D}_{S}$. In particular, let $\delta={\min}_{j\in \mathcal{D}_{S}}\omega_j - \textnormal{max}_{j\in \mathcal{I}_{S}}\omega_j$, we have the ranking consistency:
\begin{equation}
P\Bigg\{\underset{j\in \mathcal{I}_{S}}{\max} \widehat{\omega}_{j}<\underset{j\in \mathcal{D}_{S}}{\min} \widehat{\omega}_{j} \Bigg\}\ge1-2O((p-d_1)[\exp(-c'_{1}\delta^2n^{1-2\gamma})+n\exp(-c_{2}n^{\gamma})])
\end{equation}
{for a positive constant $c_1'$.}
\end{theorem}

\begin{remark}

The above theorem shows the sure screening property holds for any given conditional set $\boldsymbol{X}_S$. Define $\mathcal{M}:=\{j:\mathbb{E}(Y|X_{j})\ \text{depends on}\ X_{j}\}$ as the set of marginally dependent/active predictors. For the special case where the conditional set $\boldsymbol{X}_S=\boldsymbol{X}$, we have $\mathcal{D}_{S}=\mathcal{D}$, which is the common sure screening property in the literature. In Appendix \ref{mdc}, we also show that, similarly to Theorem \ref{Thm:sure}, the sure screening property holds when the conditional set is empty. In that case, $\textnormal{\ensuremath{\cmdc}}$ reduces to $\textnormal{\ensuremath{\mdc}}$, and the sure screening property holds for selecting the set $\mathcal{D}$ as well as $\mathcal{M}$. We discuss more details of selecting the conditional set $\boldsymbol{X}_S$ in Section \ref{subsec:S-screen}.
\end{remark}

\begin{remark}
The error terms $\exp(-c_{1}n^{1-2(\kappa+\gamma)})$ and $n\exp(-c_{2}n^{\gamma})$ in (3.5) comes from estimating the three terms in CMD$_{\mathcal{H}}$ as in Theorem 1(a). In the proof, take the first term \sloppy $E(VV'k_{1}(\boldsymbol{U}_{1},\boldsymbol{U}_{1}^{'})k_{2}(\boldsymbol{U}_{2},\boldsymbol{U}_{2}^{'})):=E(h)$ for example, we decompose it into a bounded term $E[hI(h<M)]$ plus an unbounded term $E[hI(h>M)]$ for some large enough $M>0$. Similar decomposition is done to the other two terms. Setting $M=n^\gamma$ for some $0<\gamma<1/2-\kappa$, we obtain the two error terms $\exp(-c_{1}n^{1-2(\kappa+\gamma)})$ and $n\exp(-c_{2}n^{\gamma})$ for estimating the sum of the bounded terms and that of the unbounded terms, respectively. The role of the parameter $\gamma$ is a trade-off of estimating the bounded and unbounded terms. By setting $\gamma=\frac{1-2\kappa}{3}$, we achieve a balance and obtain the optimal convergence rate. As mentioned in \citet{shao2014martingale}, their bound (3.5) can be further improved by assuming a stronger moment condition on $Y$, i.e., $E(\exp (sY^4))<\infty$ for all $s\in (0,2s_0]$. Their improved bound is the same as our bound. It is also worth mentioning that we do not impose moment conditions on the variable X as in \citet{shao2014martingale}. The reason why our method enjoys a better rate under a weaker condition is that our proposed measure CMC$_\mathcal{H}$ only computes the RKHS kernel functions of X (as shown in Theorem 1 (a)). Such kernels are bounded, which frees us from assuming additional moment conditions on X. In contrast, the Martingale Difference Correlation requires to calculate the Euclidean distance (Theorem 1 (1) in \citet{shao2014martingale}), which is unbounded. Thus, they require the stronger assumptions.
\end{remark}

\subsection{A Variable Screening Algorithm $\protect\scmdc$ \label{subsec:S-screen}}

The sure screening property holds for any conditional set. In practice,
if the conditional set is not given, we use the top $d_{1}$ predictors
suggested by $\mdc$, which also enjoys the sure screening
property, as shown in Theorems \ref{theorem: mdc-SIS} and \ref{theorem: mdc-SISQ} in the appendix. We propose to use the following variable screening algorithm $\scmdc$ as stated in Algorithm \ref{alg:scmdd}. 
\begin{algorithm}[H]
\begin{description}
\item [{{\footnotesize{}Input:}}] {\footnotesize{}The conditional set $\boldsymbol{X}_{S}$
(optional) and its cardinality $d_{1}$ (optional), the number of variables
$d_2$ to select (optional), and the data $\{(y_{i},\boldsymbol{x}_{i})\}_{i=1}^{n}$. }{\footnotesize\par}
\end{description}
\begin{enumerate}
\item {\footnotesize{}If $\boldsymbol{X}_{S}$ is not given, calculate $\mdcn(X_{i}):=\mdcn(Y,X_{i})$
for each $i=1,...,p$. Let the conditional set $\boldsymbol{X}_{S}$
be the set of the top $d_{1}$(if not given, $d_1=\left\lfloor \sqrt{n/\log n}\right\rfloor $ where $\lfloor\cdot\rfloor$
is the floor function)
predictors with the largest $\mdcn(X_{i})$.}{\footnotesize\par}
\item {\footnotesize{}For each $i\in \{1,...,p-d_1\}$, calculate $\cmdcn(X_{c_i}):=\cmdcn(Y,X_{c_i}^{\perp}|\boldsymbol{X}_{S})$,
where each $c_i$ is from ${S^c}$ and $X_{c_i}^{\perp}=X_{c_i}-P_{\boldsymbol{X}_{S}}X_{c_i}$ with $P_{\boldsymbol{X}_{S}}$
being the projection matrix onto the column space of $\boldsymbol{X}_{S}$.}{\footnotesize\par}
\item {\footnotesize{}Calculate the score $A_{i}:=\max(\frac{\mdc(X_{c_i})}{\underset{1\leq i\leq p-d_1}{\max}\mdc(X_{c_i})},\frac{\cmdc(X_{c_i})}{\underset{1\leq i\leq p-d_1}{\max}\cmdc(X_{c_i})})$
for each $i=1,...,p-d_1$. Keep the top $d_2-d_1$
predictors with the largest scores. If $d_2$ is not given, $d_2=\left\lfloor n/\log n\right\rfloor$.}{\footnotesize\par}
\end{enumerate}
\begin{description}
\item [{{\footnotesize{}Output:}}] {\footnotesize{}The index set of the
$d_2$ selected variables $\{i_{1},...,i_{d_2}\}\subseteq\{1,...,p\}$ (the $d_1$ variables selected in Step 1 plus the $d_2-d_1$ variables selected in Step 3).}{\footnotesize\par}
\end{description}
\caption{{\footnotesize{}The procedure of the $\protect\scmdc$ for variable
screening.}\label{alg:scmdd}}
\end{algorithm}
The R code of $\protect\scmdc$ is available in the supplement file. 
\begin{remark} The parameters $d_{1}$ and $d_{2}$ are predefined values. In
general, larger $d_{1}$ will lead to worse performance if the set
$\boldsymbol{X}_{S}$ contains larger proportion of
inactive variables. It will not affect the theoretical performance
of $\textnormal{\ensuremath{\cmdc}}$. However, computationally, the estimation of the expected kernel function of long
vectors inside $\textnormal{\ensuremath{\cmdc}}$ becomes less reliable if the sample size is limited.
Thus we recommend to choose small $d_{1}$(e.g., $d_1=\left\lfloor \sqrt{n/\log n}\right\rfloor$) in practice. \end{remark}

\begin{remark} The adoption of $X_{c_i}^{\perp}$ is to handle  the scenario when the independence assumption of $\boldsymbol{X}_S$ and $X_i$  is violated. Note that this only removes the linear dependence. See more discussion in Remark \ref{rmk5}.\end{remark} 

\begin{remark} In the last step, an alternative way is to use a linear
combination \sloppy $w_{1}\frac{\textnormal{\ensuremath{\mdc}}(X_{c_i})}{\underset{1\leq i\leq p-d_1}{\max}\textnormal{\ensuremath{\mdc}}(X_{c_i})}+w_{2}\frac{\textnormal{\ensuremath{\cmdc}}(X_{c_i})}{\underset{1\leq i\leq p-d_1}{\max}\textnormal{\ensuremath{\cmdc}}(X_{c_i})}$
to rank all the predictors. The weights can be adaptively chosen driven by
the data, which we leave as a potential future work. \end{remark} 

\subsection{Extension to Quantile Screening}
In this section, we extend our method to the quantile screening setting. For a univariate random response $Y$, denote $\myw =\tau-\boldsymbol{1}(Y\le q_{\tau})$ as its binary version with $\tau\in (0,1)$, where $q_{\tau}$ is the $\tau$-th quantile of the distribution of $Y$. Given $i.i.d.$ observations $\{y_k\}_{k=1}^n$ of $Y$, denote $\hatw =\tau-\boldsymbol{1}(Y\le \hat{q}_{\tau})$ as the estimate of $\myw$, where $\hat{q}_{\tau}$ is the sample $\tau$-th quantile. So for each observation $y_k$, we denote  $\mywk=\tau-\boldsymbol{1}(y_{k}\le \hat{q}_{\tau})$. Let $\omega_{j}(\myw)=\cmdc^2(\myw,X_j|\boldsymbol{X}_{S})$
and $\widehat{\omega}_{j}(\hat{Y}_\tau)=\cmdcn^2(\hatw,X_j|\boldsymbol{X}_{S})$. Similarly, we denote $\mathcal{D}_{q_{\tau}}=\{j:\mathbb{E}(Y_\tau|(\boldsymbol{X}_{S},X_j))\ \text{depends on}\ X_{j}\}$ as the quantile active predictors conditional on $\boldsymbol{X}_{S}$, and denote  $\mathcal{\widehat{D}}_{q_{\tau}}=\{j:\widehat{\omega}_{j}(\hatw)\ge cn^{-\kappa},$
for $j\in S^c\}$ as the selected variables.

Next we show the sure screening property for the quantile version
of $\cmdc$.
\begin{theorem} Under condition $(C1)$ in the appendix, for any $0<\gamma<1/2-\kappa$
and $\kappa\in(0,1/2)$, there exist positive constants $c_{1},c_{2}$
such that for any $c>0$, 
\begin{equation}
P\Bigg\{\underset{1\leq j\leq p-d_1}{\max}|\widehat{\omega}_{j}(\hat{Y}_\tau)-\omega_{j}(\myw)|\ge cn^{-\kappa}\Bigg\}\le O((p-d_1)[\exp\{-c_{1}n^{1-2(\kappa+\gamma)}\}+n\exp(-c_{2}n^{\gamma})]).
\end{equation}
If Condition $(C2)$ in the appendix holds in addition, we have 
\begin{equation}
P(\mathcal{D}_{q_{\tau}}\subseteq\hat{\mathcal{D}}_{q_{\tau}})\ge1-O(\widetilde{s}_{n}[\exp\{-c_{1}n^{1-2(\kappa+\gamma)}\}+n\exp(-c_{2}n^{\gamma})]),
\end{equation}
where $\widetilde{s_{n}}$ is the cardinality of $\mathcal{D}_{q_{\tau}}$.

\end{theorem}


\section{Numerical Studies\label{section: simulation}}

In this section, we evaluate the finite-sample performance of the
proposed method $\cmdc$. 
\paragraph{The choice of kernel functions}
Denote the translation-invariant Gaussian kernel as 
\begin{align}
K(\boldsymbol{x}_{1},\boldsymbol{x}_{2}):=\exp(-\frac{1}{h}(\boldsymbol{x}_{1}-\boldsymbol{x}_{2})^{T}(\boldsymbol{x}_{1}-\boldsymbol{x}_{2})),
\end{align}
where $\boldsymbol{x}_{1},\boldsymbol{x}_{2}\in\mathbb{R}^{t}, t\in\mathbb{N}$, and $h$ is the bandwidth. For the proposed $\cmdc$,
we use $K(\boldsymbol{x}_{1},\boldsymbol{x}_{2})$ for both $k_{1}$
and $k_{2}$. In our simulations, the performance of $\cmdc$ for
variable screening is robust against the bandwidths of $k_{1}$ and
$k_{2}$. So we set $h=2$ for both $k_{1}$ and $k_{2}$. For $\mdc$,
we adopt Gaussian kernel and conduct a sensitivity analysis of the bandwidth, {the results of which are presented in the supplementary material S1.1. The performance of $\mdc$ is sensitive
to the bandwidth $h$. In particular, $\mdc$  with smaller $h$ performs better for selecting covariates that are linearly related to the response variable, while
larger $h$ is more suitable for selecting nonlinearly related covariates. To select the conditional set $\boldsymbol{X}_S$ and avoid cherry-picking, we first calculate the
values of $\mdc$ using two bandwidths $h=2\hat{\sigma}_{X_i}^{2}$
and $h=6\hat{\sigma}_{X_i}^{2}$ for each predictor $X_i$, where $\hat{\sigma}_{X_i}^{2}$ is sample
variance of $X_i$ and $i=1,...,p$. Our experience is that $h=2\hat{\sigma}_{X_i}^{2}$ and $h=6\hat{\sigma}_{X_i}^{2}$ generally perform well for the simulations. Then we take the maximum of the two $\mdc$ values for each predictor. One can also use Laplacian kernel
and Cauchy kernel in practice. However, in our examples, they yield
similar performance to that of the Gaussian kernel. 

\paragraph{Criteria of evaluating variable screening performance}

Following \citet{li2012feature}, we consider three criteria for evaluating
the variable screening performance: 1) $\mathcal{S}_{q}$: the $(100q)$-th
quantile of the minimal model size required to contain all the active predictors. 2) $\mathcal{P}_{i}$: the
proportion that the predictor $X_{i}$ is selected, and 3) $\mathcal{P}_{\textnormal{all}}$:
the proportion of all active predictors being selected. Essentially,
an $\mathcal{S}_{q}$ closer to the total number of active predictors
is preferred. The three criteria are connected in a way that a smaller
minimal model size suggests a larger $\mathcal{P}_{all}$ and a lager
$\mathcal{P}_{i}$ for each 
active variable.

\paragraph{Screening thresholds}

We compare $\scmdc$ with six variable screening methods,
including two marginal screening method (DCSIS2 in \citet{kong2017interaction} and MDC), three conditional screening
methods (CSIS in \citet{barut2016conditional}, CDCSIS in \citet{wen2018sure}  and CIS in \cite{tong2022model}), and one iterative methods
(RaSE$_{1}$-eBIC in \citet{tian2021rase}.
For each screening method, we keep the top $d_2=\lfloor n/log(n)\rfloor$
variables, where $n$ is the sample size. We report the $\mathcal{P}_{i}$, $\mathcal{P}_{all}$
and $\mathcal{S}_{0.5}$ values based on 100 repetitions for each example. Since conditional screening methods require a
pre-selected conditional set $\boldsymbol{X}_{S}$, we either artificially
set up the conditional set based on the variables in the true model
or select the top $d_{1}=\lfloor\sqrt{n/log(n)}\rfloor$ variables
suggested by $\mdc$. We also include a sensitivity analysis of using different methods (e.g. SIS, LASSO and forward regression) to choose the conditional set in the supplementary material S1.2.

\subsection{Simulation}

\paragraph{Example 1 (Marginally inactive but jointly active predictors).}

Following the idea of \citet{fan2008sure}, we generate samples $\{Y_{i},\boldsymbol{X}_{i}\}_{i=1}^{n}$
from the linear regression model
\[
Y=X_{1}+X_{2}+X_{3}+X_{4}+X_{5}-cX_{6}+\epsilon,
\]
where the coefficient $c$ is designed so that $\textrm{cov}(X_{6},Y)=0$.
That is, the predictor $X_{6}$ is marginally independent from the
response $Y$. 
 The predictor vector $\boldsymbol{X}=(X_{1},...,X_{p})\sim N(\boldsymbol{0},\boldsymbol{\Sigma}$),
where $\boldsymbol{\Sigma}$ has $(i,j)$-th entry $\sigma_{ij}=\rho^{I\{i\neq j\}}$.
The error term $\epsilon\sim N(0,1)$ and is independent from $\boldsymbol{X}$.
We set the sample size $n=200$ and the dimension $p=3000$. We consider
three cases: $(c,\rho)\in\{(2.5,0.5),(4,0.8),(4.5,0.9)\}$. Note that
$X_{6}$ is dependent with $Y$ if given one or more predictors from
$\{X_{1},...,X_{5}\}$. 

\begin{table}
\centering{}%
\scalebox{0.61}{\begin{tabular}{lllllllll}
\hline 
 & $\mathcal{P}_{1}$  & $\mathcal{P}_{2}$  & $\mathcal{P}_{3}$  & $\mathcal{P}_{4}$  & $\mathcal{P}_{5}$  & $\mathcal{P}_{6}$  & $\mathcal{P}_{all}$  & $\mathcal{S}_{0.5}$ \tabularnewline
\hline 
\multicolumn{9}{c}{ $c=2.5$, $\rho=0.5$}\tabularnewline
\hline 
MDC  & 0.89  & 0.94  & 0.93  & 0.91  & 0.90  & 0.00  & 0.00  & 3000.0 \tabularnewline
CSIS ($\boldsymbol{X}_{S_{1}}$) & 1.00  & 0.97  & 0.98  & 0.98  & 0.97  & 0.87  & 0.78  & 14.0 \tabularnewline
CSIS ($\boldsymbol{X}_{S_{2}}$) & 0.71 & 0.75  & 0.75  & 0.72  & 0.72 & 1.00  & 0.17  & 1580.5 \tabularnewline
CDC-SIS ($\boldsymbol{X}_{S_{1}}$) & 1.00  & 0.95  & 0.95  & 0.95  & 0.93  & 0.16  & 0.14  & 708.5 \tabularnewline
CDC-SIS ($\boldsymbol{X}_{S_{2}}$) & 0.73  & 0.76  & 0.74  & 0.72  & 0.72  & 1.00  & 0.16  & 1541.5 \tabularnewline
CIS($\boldsymbol{X}_{S_{1}}$)& 1.00 & 0.83 & 0.88 & 0.87  & 0.87 & 0.71  & 0.34  & 58.0\tabularnewline
CIS($\boldsymbol{X}_{S_{2}}$)& 0.97 & 0.91 & 0.94 & 0.94  & 0.97 & 0.01  & 0.01  & 2997.5\tabularnewline
$\scmdc$ ($\boldsymbol{X}_{S_{1}}$) & 1.00& 0.94& 0.94 &0.92& 0.92& 0.53&    0.39     &  90.5 \tabularnewline
$\scmdc$ ($\boldsymbol{X}_{S_{2}}$) & 0.91 & 0.95 & 0.94 & 0.93 & 0.92  & 1.00 & 0.72  & 17.5 \tabularnewline
RaSE$_{1}$-eBIC & 0.99  & 1.00 & 0.99 & 0.99 & 1.00 & 1.00 & 0.97 & 6.0\tabularnewline

\hline 
\multicolumn{9}{c}{$c=4$, $\rho=0.8$}\tabularnewline
\hline 
MDC  & 0.61  & 0.63  & 0.64  & 0.63  & 0.62  & 0.00  & 0.00  & 3000.0 \tabularnewline
CSIS ($\boldsymbol{X}_{S_{1}}$) & 1.00  & 0.48  & 0.44  & 0.40  & 0.44  & 1.00  & 0.10  & 320.0 \tabularnewline
CSIS ($\boldsymbol{X}_{S_{2}}$) & 0.39  & 0.34  & 0.36  & 0.41  & 0.33  & 1.00  & 0.00  & 2988.5 \tabularnewline
CDC-SIS ($\boldsymbol{X}_{S_{1}}$) & 1.00  & 0.56  & 0.53  & 0.54  & 0.54  & 0.93  & 0.06  & 264.0 \tabularnewline
CDC-SIS ($\boldsymbol{X}_{S_{2}}$) & 0.39  & 0.34  & 0.36  & 0.41  & 0.34  & 1.00  & 0.00  & 2623.5 \tabularnewline
CIS($\boldsymbol{X}_{S_{1}}$)& 1.00 & 0.39 & 0.31 & 0.34  & 0.42 & 1.00  & 0.03  & 546.0\tabularnewline
CIS($\boldsymbol{X}_{S_{2}}$)& 0.70 & 0.59 & 0.67 & 0.64  & 0.69 & 1.00  & 0.09  & 268.0\tabularnewline
$\scmdc$ ($\boldsymbol{X}_{S_{1}}$) &  1.00 & 0.73 & 0.71  &0.66 & 0.67  &1.00  &   0.26   &    127.5 \tabularnewline
$\scmdc$ ($\boldsymbol{X}_{S_{2}}$) & 0.61  & 0.73  & 0.72  & 0.66 & 0.67  & 1.00  & 0.16 & 156.5 \tabularnewline
RaSE$_{1}$-eBIC & 0.80 & 0.92  & 0.76 & 0.79 & 0.87 & 1.00 & 0.35  & 2312.5\tabularnewline

\hline 
\multicolumn{9}{c}{$c=4.5$, $\rho=0.9$}\tabularnewline
\hline 
MDC  & 0.48  & 0.40  & 0.39  & 0.48  & 0.40  & 0.00  & 0.00  & 3000.0 \tabularnewline
CSIS ($\boldsymbol{X}_{S_{1}}$) & 1.00  & 0.14  & 0.09  & 0.13  & 0.08  & 1.00  & 0.00  & 2339.0 \tabularnewline
CSIS ($\boldsymbol{X}_{S_{2}}$) & 0.28  & 0.24  & 0.23 & 0.24 & 0.20  & 1.00  & 0.00  & 2992.5 \tabularnewline
CDC-SIS ($\boldsymbol{X}_{S_{1}}$) & 1.00  & 0.29  & 0.28  & 0.29  & 0.28  & 0.98  & 0.01  & 875.0 \tabularnewline
CDC-SIS ($\boldsymbol{X}_{S_{2}}$) & 0.28  & 0.24 & 0.23 & 0.24 & 0.21 & 1.00  & 0.00  & 2875.0 \tabularnewline
CIS($\boldsymbol{X}_{S_{1}}$)& 1.00 & 0.11 & 0.08 & 0.07  & 0.16 & 1.00  & 0.00  & 2037.5\tabularnewline
CIS($\boldsymbol{X}_{S_{2}}$)& 0.43 & 0.44 & 0.47 & 0.46  & 0.41 & 1.00  & 0.00  & 897.0\tabularnewline
$\scmdc$ ($\boldsymbol{X}_{S_{1}}$) & 1.00&  0.48&  0.47 & 0.51&  0.45 & 1.00 &    0.07   &    269.5\tabularnewline
$\scmdc$ ($\boldsymbol{X}_{S_{2}}$) & 0.53 & 0.48  & 0.47 & 0.51 & 0.47  & 1.00  & 0.03  & 338.0 \tabularnewline
RaSE$_{1}$-eBIC & 0.62 & 0.89 & 0.73 & 0.65  & 0.68 & 1.00  & 0.19  & 2316.0\tabularnewline

\hline 
\end{tabular}}\caption{The $\mathcal{P}_{i}$, $\mathcal{P}_{all}$ and $\mathcal{S}_{0.5}$
in Example 1. The conditional set is either $\boldsymbol{X}_{S_{1}}=\{X_{1}\}$
or $\boldsymbol{X}_{S_{2}}=\{\textrm{the first \ensuremath{d_{1}=6} predictors selected by MDC\ensuremath{_{\mathcal{H}}}}\}$.\label{table: eg1-CS} }
\end{table}

The simulation results are reported in Table \ref{table: eg1-CS}.
The marginal screening method MDC fails to detect the marginally independent
predictor $X_{6}$ in all cases, while other methods identify  $X_{6}$ as an active predictor. As the correlation increases from $\rho=0.5$ to $\rho=0.9$, the
selection proportion $\mathcal{P}_{i}$ decreases for $i=1,2,...,5$.
Consequently, $\mathcal{P}_{all}$ decreases as the correlation increases. Note that if the conditional set $\boldsymbol{X}_{S_{1}}=\{X_{1}\}$,
the proportion of selecting $X_{1}$ is set to 1. We set $d_2=\left\lfloor {n/log(n)}\right\rfloor =37$ in this example.
Note that the minimal model size $\mathcal{S}_{0.5}$ is small only in the first case where $\rho=0.5$, which explains the low
values of $\mathcal{P}_{all}$ for all the methods for the high correlation
case ($\rho=0.8$ or $0.9$). The three conditional methods: CIS, CSIS and
CDC-SIS, fail to detect the active variables $X_{2},X_{3},X_{4}$
and $X_{5}$ when the correlation increases to $\rho=0.9$.

When the conditional set changes from an oracle set $\boldsymbol{X}_{S_{1}}$
to a data-dependent set $\boldsymbol{X}_{S_{2}}$, the performances
of the three conditional screening methods (CIS, CSIS and CDC-SIS) deteriorate.
Our proposed S-C$\mdc$ instead shows robustness against the conditional
set. The performance of our method decreases due to the extra cost
of selecting $X_{1}$ when the conditional set becomes data-dependent. 

In this example, from the perspective of the minimal model
size $\mathcal{S}_{0.5}$, the minimal model size for RaSE$_{1}$-eBIC
changes from $6$ to $2316$ when $\rho$ increases from $0.5$
to $0.9$. In contrast, the stable performance of S-$\cmdc$ in $\mathcal{S}_{0.5}$ indicates that our method is more robust against the correlation $\rho$. In terms of $\mathcal{P}_{all}$ and $\mathcal{P}_{i}$, RaSE$_{1}$-eBIC and S-$\cmdc$ are better than any other method. RaSE$_{1}$-eBIC is better than S-$\cmdc$, which may be due to the good performance of RaSE$_{1}$-eBIC in the linear case.

\paragraph{Example 2 (Interaction terms).}

We consider the following model with interaction terms: 
\begin{align*}
Y= & X_{1}+X_{5}+X_{10}+X_{1}X_{15}+1.5X_{5}X_{20}+2X_{10}X_{25}+\epsilon.
\end{align*}
The predictor vector $\boldsymbol{X}=(X_{1},...,X_{p})\sim N(\boldsymbol{0},\boldsymbol{\Sigma}$),
where $\boldsymbol{\Sigma}$ has $(i,j)$-th entry $\sigma_{ij}=\rho^{|i-j|}$.
We consider two cases: $\rho\in\{0,0.9\}$. The error term $\epsilon\sim N(0,1)$
and is independent from $\boldsymbol{X}$. We set the sample size
$n=200$ and the dimension $p=3000$. 

\begin{table}
\centering{}%
\scalebox{0.8}{\begin{tabular}{lllllllll}
\hline 
 & $\mathcal{P}_{1}$ & $\mathcal{P}_{5}$ & $\mathcal{P}_{10}$ & $\mathcal{P}_{15}$ & $\mathcal{P}_{20}$ & $\mathcal{P}_{25}$ & $\mathcal{P}_{all}$ & $\mathcal{S}_{0.5}$\tabularnewline
\hline 
\multicolumn{9}{c}{$\rho=0$}\tabularnewline
\hline 
MDC & 0.94 & 0.96 & 0.90 & 0.01 & 0.00 & 0.03 & 0.00 & 2662.5\tabularnewline
CSIS ($\boldsymbol{X}_{S_{1}}$) & 1.00 & 1.00 & 1.00 & 0.04 & 0.03 & 0.07 & 0.00 & 2452.0\tabularnewline
CSIS($\boldsymbol{X}_{S_{2}}$) & 0.97 & 0.93 & 0.90 & 0.03 & 0.02 & 0.04 & 0.00 & 2147.0\tabularnewline
CDC-SIS ($\boldsymbol{X}_{S_{1}}$) & 1.00 & 1.00 & 1.00 & 0.01 & 0.03 & 0.26 & 0.00 & 1639.0\tabularnewline
CDC-SIS ($\boldsymbol{X}_{S_{2}}$)& 0.84 & 0.86 & 0.84 & 0.07 & 0.17 & 0.43 & 0.00 & 1529.0\tabularnewline
DCSIS2 & 0.40 & 0.70 & 0.99 & 0.11 & 0.35 & 0.85 & 0.00 & 1372.5\tabularnewline
CIS($\boldsymbol{X}_{S_{1}}$) & 1.00 & 1.00 & 1.00 & 0.02 & 0.12 & 0.45 & 0.00 & 566.0\tabularnewline
CIS($\boldsymbol{X}_{S_{2}}$) & 0.99 & 0.98 & 0.96 & 0.01 & 0.01 & 0.02 & 0.00 & 2278.0\tabularnewline
$\scmdc$ ($\boldsymbol{X}_{S_{1}}$) & 1.00 & 1.00 & 1.00 & 0.18 & 0.62 & 0.99 & 0.09 & 196.5\tabularnewline
$\scmdc$($\boldsymbol{X}_{S_{2}}$) & 0.94 & 0.96 & 0.91 & 0.11 & 0.37 & 0.77 & 0.01 & 851.5\tabularnewline
RaSE$_{1}$-eBIC & 0.87 & 0.82 & 0.79 & 0.01 & 0.01 & 0.04 & 0.00 & 2295.5\tabularnewline

\hline 
\multicolumn{9}{c}{$\rho=0.9$}\tabularnewline
\hline 
MDC &1.00 & 1.00 & 1.00 & 0.97 & 0.52 & 0.14   &  0.13    &   350.5\tabularnewline
CSIS ($\boldsymbol{X}_{S_{1}}$) &1.00  &1.00 & 1.00 &0.08 &0.15 &0.13  &  0.02  &   1887.0\tabularnewline
CSIS ($\boldsymbol{X}_{S_{2}}$)& 0.78& 0.97 & 0.74 & 0.08 & 0.12 & 0.13 &    0.02  &    1870.0\tabularnewline
CDCSIS ($\boldsymbol{X}_{S_{1}}$) & 1.00&   1.00&   1.00 & 0.03&  0.04&  0.06 &    0.00 &     2024.5\tabularnewline
CDCSIS($\boldsymbol{X}_{S_{2}}$) & 0.52& 0.96 &0.80 &0.86& 1.00 &1.00   & 0.40      & 66.0\tabularnewline
DCSIS2 & 0.93  &1.00 & 1.00 &0.87 &0.91&0.81 &   0.62  &     30.5\tabularnewline
CIS($\boldsymbol{X}_{S_{1}}$) & 1.00  &   1.00&  1.00 &0.28& 0.93& 1.00 &      0.27 &        62.0\tabularnewline
CIS($\boldsymbol{X}_{S_{2}}$) &  0.78 & 0.99&  0.83&  0.43&  0.92 & 0.85 &     0.2  &      88.5\tabularnewline
$\scmdc$($\boldsymbol{X}_{S_{1}}$) & 1.00&   1.00&   1.00&  0.90&  1.00&  1.00&    0.90 &       25.0\tabularnewline
$\scmdc$($\boldsymbol{X}_{S_{2}}$) & 1.00 & 1.00&  1.00&  0.90&  1.00&  0.99 &    0.89 &       26.0\tabularnewline
RaSE$_{1}$-eBIC & 0.41& 0.42& 0.37& 0.02 &0.02 &0.01 &   0.01    & 2403.0\tabularnewline

\hline 
\end{tabular}}\caption{The $\mathcal{P}_{i}$, $\mathcal{P}_{all}$ and $\mathcal{S}_{0.5}$
in Example 2.
\label{table: eg2-NCS}We set the conditional set to be $\boldsymbol{X}_{S_{1}}$=$\{X_{1},X_{5},X_{10}\}$
or $\boldsymbol{X}_{S_{2}}=$\{the first $d_{1}=6$ predictors selected
by $\protect\mdc$\}.}
\end{table}

The results are reported in Table \ref{table: eg2-NCS}. The three
variables $X_{15},X_{20},X_{25}$ all jointly contribute to the mean
of $Y$, but are marginally independent of the mean of $Y$. It is difficult for the marginal screening methods  to detect these three terms. Note that $X_{25}$ has a larger coefficient in its interaction
term than that of $X_{20}$ or $X_{15}$. As the signal/coefficient
of the interaction term increases, its effect is easier to
be detected ($\mathcal{P}_{25}>\mathcal{P}_{20}>\mathcal{P}_{15}$) for all the methods. RaSE$_{1}$-eBIC fails in detecting
those three variables in this example. This is possibly due to the fact that
RaSE$_{1}$-eBIC targets additive models. In comparison to the interaction
screening method DCSIS2, the proposed method S-C$\mdc$ has a comparable
performance in selecting $X_{15}$, $X_{20}$ and $X_{25}$ for both $\rho=0$ and
$\rho=0.9$. But S-C$\mdc$ performs better than DCSIS2 in selecting
the marginally active variables $X_{1}$, $X_{5}$ and $X_{10}$. What's more, S-$\cmdc$ has the smallest $\mathcal{S}_{0.5}$ among all the methods. Similar to Example 1, the performance of S-$\cmdc$ is stable against
the conditional set. We also did a sensitivity analysis against the conditional set. We consider three more cases: the conditional set is selected by Lasso, SIS, and forward regression. The results are reported in the supplementary material S1.2, demonstrating a better and more stable performance of $\scmdc$ compared to other conditional methods (CSIS, CDC-SIS). Finally, we include a block structure correlation setting where the correlation among active predictors are 0.2 and 0.1 otherwise. We report the result in the supplementary material S1.3, which clearly shows the advantage of S-$\cmdc$. 

\paragraph{Example 3 (Heteroscedasticity \& Quantile screening).}

In this example, we demonstrate that our screening method can help
in heteroskedastic model specification. We consider the following
model: 
\begin{align*}
Y= & X_{1}+X_{5}+X_{1}X_{10}+1.5X_{5}X_{15}+\epsilon\cdot\exp(X_{35}+X_{40}).
\end{align*}
The predictor vector $\boldsymbol{X}=(X_{1},...,X_{p})\sim N(\boldsymbol{0},\boldsymbol{\Sigma}$),
where $\boldsymbol{\Sigma}$ has $(i,j)$-th entry $\sigma_{ij}=\rho^{|i-j|}$.
We consider two cases: $\rho\in\{0,0.9\}$. The error term $\epsilon\sim N(0,1)$
is independent from $\boldsymbol{X}$. We set the sample size $n=400$
and the dimension $p=3000$. For the purpose of quantile screening,
we change the continuous response $Y$ to a binary response $Y_{\textrm{\ensuremath{\tau}}}=\tau-\boldsymbol{1}(Y\le\hat{q}_{\tau})$,
where $\hat{q}_{\tau}$ is the $\tau$-th sample quantile of the response.
Then we apply S-$\cmdc$ and MDC
on the data $(Y_{\ensuremath{\tau}},\boldsymbol{X})$.
Note that in fixed design, the population quantile $q_{\tau}$ does
not depend on $(X_{20},X_{25})$ if and only if $\tau=0.5$. We consider
two choices of the quantile: $\tau=0.5$ and $\tau=0.75$. 

The results are presented in Table \ref{table: eg3-NCS} with conditional set selected by $\protect\mdc$. In this example, we also evaluate the method QaSIS \citep{he2013quantile}, a quantile-adaptive model-free variable screening method for heterogeneous data. Overall,
our proposed method S-C$\mdc$ performs the best across all four combinations of $(\rho,\tau)$. In particular, S-C$\mdc$ is dominantly better in ${\cal S}_{0.5}$ than any other method even under high correlation setting. When there is no correlation among predictors ($\rho=0$), from the perspective of quantile screening,
MDC, CDC-SIS, QaSIS, and S-C$\mdc$ correctly differentiate the different
roles of $X_{35}$ and $X_{40}$ under two values of $\tau$. However,
these three methods (MDC, CDC-SIS, QaSIS) fail to identify $X_{15}$ (compared to $X_{10}$) when $\tau=0.5$. In contrast, S-C$\mdc$ has a better performance in separating the active variables from the inactive variables. When high correlation exists among predictors, all methods receive improved performance and S-C$\mdc$ still remains competitive. This is because the marginal relationship between $X_{10}$ ($X_{15}$) and the response $Y$ is strengthened by the high correlation among predictors. We also include the scenario when conditional set is $\{X_{1},X_{5},X_{10}\}$ in the supplementary material S1.4. 

\begin{table}
\begin{centering}
\footnotesize
\scalebox{0.8}{\begin{tabular}{lclllllllllllllllll}
\hline 
\multirow{2}{*}{Method} & \multirow{2}{*}{$\tau$} & \multicolumn{8}{c}{$\rho=0$} &  & \multicolumn{8}{c}{$\rho=0.9$}\tabularnewline
\cline{3-10} \cline{4-10} \cline{5-10} \cline{6-10} \cline{7-10} \cline{8-10} \cline{9-10} \cline{10-10} \cline{12-19} \cline{13-19} \cline{14-19} \cline{15-19} \cline{16-19} \cline{17-19} \cline{18-19} \cline{19-19} 
 &  & $\mathcal{P}_{1}$ & $\mathcal{P}_{5}$ & $\mathcal{P}_{10}$ & $\mathcal{P}_{15}$ & $\mathcal{P}_{35}$ & $\mathcal{P}_{40}$ & $\mathcal{P}_{all}$ & $\mathcal{S}_{0.5}$ &  & $\mathcal{P}_{1}$ & $\mathcal{P}_{5}$ & $\mathcal{P}_{10}$ & $\mathcal{P}_{15}$ & $\mathcal{P}_{35}$ & $\mathcal{P}_{40}$ & $\mathcal{P}_{all}$ & $\mathcal{S}_{0.5}$\tabularnewline
\hline 
\multirow{2}{*}{MDC } & 0.5 & 1.00 & 1.00 & 0.06 & 0.03 & 0.04  &0.02  &   0.00  & 1910.0 &  & 1.00 &1.00& 0.97 &0.49 &0.06& 0.05 &   0.49  &     71.0 \tabularnewline
 &0.75&  0.99 & 1.00&  0.24 & 0.38&  0.52&  0.55&     0.03   &    986.5 &  &  1.00 &1.00& 1.00 &1.00 &0.91 &0.92  &  0.86 &      29.5\tabularnewline
\cline{1-2} \cline{2-2} 
\multirow{2}{*}{CSIS}  & 0.5 & 1.00& 0.99& 0.01 &0.03& 0.20 &0.16  &  0.00  &   2025.5&  &  1.00 &  1.00&   0.19&   0.07 &  0.31&   0.24 &     0.00    &   1854.5\tabularnewline
 & 0.75 & 0.97&  0.95 & 0.06&  0.14&  0.39 & 0.42 &    0.00    &  2207.0 &  &  0.86&  0.96&  0.39& 0.11 & 0.31&  0.30 &    0.00 &     2138.5\tabularnewline
\cline{1-2} \cline{2-2} 
\multirow{2}{*}{CDC-SIS} & 0.5 & 1.00 & 0.99&  0.13&  0.28&  0.23&  0.30  &   0.04 &757.0 &  & 1.00 &  1.00 &  0.28 &  0.94 &  1.00  & 1.00  &   0.27    &    172.5\tabularnewline
 & 0.75 & 0.96&  0.97&  0.13&  0.51 & 0.42 & 0.47 &    0.04  &    1159.5 &  & 0.88& 0.97& 0.41& 0.87 &0.95& 0.92  &  0.25   &   214.5\tabularnewline
\cline{1-2} \cline{2-2} 
\multirow{2}{*}{QaSIS } & 0.5 & 1.00  &1.00 & 0.16 & 0.16 & 0.28 & 0.27   &  0.03  & 1085.0&  &1.00 &1.00 &1.00 &0.90 &0.69 &0.55 &   0.90&       20.5\tabularnewline
 & 0.75 &  0.92 &0.99 &0.14& 0.38 &0.69 &0.72  &  0.03  &    600.0 &  & 0.99 & 1.00 & 1.00 &0.97  &0.99 & 0.97  &   0.92    &    35.5\tabularnewline
\cline{1-2} \cline{2-2} 
\multirow{2}{*}{DCSIS2 } &0.5& 0.18& 0.41 &0.04 &0.08 &0.93& 0.90  &  0.00 &    1443.5 &  & 0.07  & 0.15  & 0.06 &  0.05  & 1.00 &  1.00  &    0.00   &     918.0\tabularnewline
 & 0.75 &0.18 &0.41& 0.04& 0.08 &0.93 &0.90  &  0.00  &   1443.5&  & 0.07 &0.15& 0.06 &0.05& 1.00& 1.00 &   0.00     & 918.0\tabularnewline

 \cline{1-2} \cline{2-2} 
 \multirow{2}{*}{CIS}&0.5&   1.00 &0.99 &0.07& 0.04 &0.01& 0.02&    0.01  &   1982.5
& & 1.00 & 1.00 &0.31& 0.89 &0.95& 0.98 &   0.29   &   121.5\tabularnewline
&0.75& 1.00 & 0.99&  0.06&  0.14&  0.21&  0.3  &      0.00  &    2537.5& 
&  1.00& 0.99 &0.47 &0.76& 0.9& 0.91 &   0.33  &      129.0\tabularnewline
\cline{1-2} \cline{2-2} 
\multirow{2}{*}{$\scmdc$} & 0.5 &1.00& 1.00& 0.18& 0.80 &0.05& 0.04  &  0.13    &  240.0 &  & 1.00& 1.00 &0.97 &0.93 &0.34 &0.38   & 0.90   &    16.0\tabularnewline
 & 0.75 &0.99  &1.00 & 0.33 & 0.67 & 0.49 & 0.58  &   0.05    &   574.5 &  &  1.00 &1.00& 1.00&0.95 &0.97& 0.93  &  0.86    &   28.0\tabularnewline
\cline{1-2} \cline{2-2} 
\multirow{2}{*}{RaSE$_{1}$-eBIC} &0.5 &0.49& 0.31& 0.00 &0.00 &0.03 &0.01 &   0.00 &1816.0&  & 0.25 &0.07 &0.00 &0.00& 0.04 &0.03  &  0.00    & 2425.5\tabularnewline
 &0.75 & 0.49&  0.31 & 0.00&  0.00 & 0.00 & 0.00 &   0.00   &   2427.0 &  & 0.25 &0.07& 0.00& 0.00& 0.01 &0.02 &   0.00&     2427.0\tabularnewline

\hline 
\end{tabular}}{\footnotesize\par}

\caption{The $\mathcal{P}_{i}$, $\mathcal{P}_{all}$ and $\mathcal{S}_{0.5}$
in Example 3 with $\boldsymbol{X}_{S}$ selected by $\protect\mdc$\label{table: eg3-NCS}.}
\par\end{centering}
\end{table}

\paragraph{Example 4 (Nonlinear case).}
We consider the following nonlinear case.
\begin{equation*}
Y=3I(X_{1}>0.5)X_{2}+3sin^2(2\pi X_{1})X_{3}+3(X_{1}^2-1)X_{4}+\exp(X_{1})X_{5}+\epsilon,
\end{equation*}
where we first generate $U_{1}$, $U_{2}$ $\overset{i.i.d}{\sim}\textrm{Unif}[0,1]$
and then let $X_{1}=(U_{1}+U_{2})/2$ and $X_{k}=(Z_{k}+2U_{1})/4$ for
$k=2,3,...,p$. We consider two scenarios, symmetric distribution $Z_{k}\stackrel{i.i.d}{\sim}N(0,1)$
and asymmetric distribution $Z_{k}\stackrel{i.i.d}{\sim}\chi^2_{(1)}$. The error term $\epsilon$ is independently drawn from standard normal distribution. We set the sample size $n=200$ and the dimension
$p=3000$. The results are reported in Table \ref{table: eg4b}.

\begin{table}
\centering{}%
\scalebox{0.7}{\begin{tabular}{llllllllllllllll}
\hline 
\multicolumn{7}{c}{$Z_{k}$ $\stackrel{i.i.d}{\sim}N(0,1)$}  &  & \multicolumn{7}{c}{$Z_{k}$ $\stackrel{i.i.d}{\sim}\chi^2_{(1)}$}\tabularnewline \hline

 & $\mathcal{P}_{1}$  & $\mathcal{P}_{2}$  & $\mathcal{P}_{3}$  & $\mathcal{P}_{4}$  & $\mathcal{P}_{5}$    & $\mathcal{P}_{all}$  & $\mathcal{S}_{0.5}$ &&   $\mathcal{P}_{1}$  & $\mathcal{P}_{2}$  & $\mathcal{P}_{3}$  & $\mathcal{P}_{4}$  & $\mathcal{P}_{5}$    & $\mathcal{P}_{all}$  & $\mathcal{S}_{0.5}$\tabularnewline

\hline 
MDC  & 1.00  & 0.79  & 0.84  & 0.00  & 0.85    & 0.00  & 2974.5 && 1.00&0.64&0.80&0.00&0.85&0.00&2725.0 \tabularnewline 
CSIS  & 0.99  & 0.56   & 0.55  & 1.00  & 0.69  & 0.17  & 554.5&& 1.00&0.61&0.70&1.00&0.81&0.26&100.5 \tabularnewline
CDC-SIS  & 0.99    & 0.62  & 0.53  & 0.34  & 0.68  & 0.11  & 1101.5&& 1.00&0.45&0.60&0.65&0.60&0.11&962.5  \tabularnewline
DCSIS2    &  1.00   &  0.84   &  0.57    & 0.00    & 0.52     &   0.00      &   2235.0   &   & 1.00   & 0.90  & 0.71  & 0.05  & 0.70  & 0.01  & 1284.5 \tabularnewline
CIS  & 1.00  & 0.95 & 0.90 & 0.31  & 0.94 & 0.24  & 153.0 &&  1.00 & 0.76 & 0.78 & 0.42  & 0.78 & 0.17  & 155.0 \tabularnewline
$\scmdc$  & 1.00  & 0.79 & 0.84 & 0.95  & 0.85 & 0.52  & 33.5 && 1.00&0.63&0.78&0.90&0.82&0.35&77.0 \tabularnewline

RaSE$_{1}$-eBIC & 0.28  & 0.60 & 0.60  & 0.58 & 0.74 & 0.06 & 2250.0&& 0.98&0.47&0.63&0.95&0.73&0.18&2253.5 \tabularnewline
\hline 
\end{tabular}}\caption{The $\mathcal{P}_{i}$, $\mathcal{P}_{all}$ and $\mathcal{S}_{0.5}$
in Example 4 with  $\boldsymbol{X}_S$ selected by  $\protect\mdc$.\label{table: eg4b}}
\end{table}

In this example, the active predictors are nonlinearly associated with each other. Each element in this model is an interaction effect of two variables. The proposed S-CMC$_\mathcal{H}$ demonstrates its competitive performance for selecting all the active predictors with relatively high $\mathcal{P}_i$ and $\mathcal{P}_{all}$, and smallest $\mathcal{S}_{0.5}$, in both scenarios. The methods DCSIS2 and MDC fail to detect $X_4$. It is worth pointing out that the contribution of variable $X_4$ is also underestimated by the methods CDC-SIS and CIS.
This may be because that $X_4$'s conditional contribution to $Y$ is diffused by the exponential term exp$(X_1) X_5$, as indicated by the much larger $\mathcal{P}_5$ in the four methods mentioned above.

\subsection{Data Applications}

\subsubsection{Single Cell Malt Tumor CITE-seq Dataset}\label{realdata1}
The Malt Tumor \textbf{C}ellular \textbf{I}ndexing of \textbf{T}ranscriptomes and \textbf{E}pitopes by sequencing (CITE-seq) dataset (\textit{http://www.10xgenomics.com}) contains single-cell level sequencing RNA data and as well as the surface protein expression count. We are interested in identifying the genes that affect the surface protein level. The dataset contains 33555 genes and proteins from 8412 single cells. Following the data pre-processing procedure in \cite{tong2022model} (filtering out cells with more than 90\% zero entries and genes that has zero variance), we obtain a sample of  $n=207$ single cells and $p=18702$ genes. And we set protein CD8 as the response variable and the protein CD3 as conditional variable. Interested readers are referred to \cite{tong2022model} for detailed scientific explanations of using CD3 as the conditional variable. To evaluate the prediction performance of each screening method, we randomly split the observations into a training set of size 176 and a test set of size 31. We select $d_2=38$ variables by the following four conditional methods: CSIS, CDCSIS, CIS and SCMC$_\mathcal{H}$ including CD3. Then a random forest model is fitted with the selected variables and the response, with log transformations on the response and the variables. The mean squared errors (MSE) of the methods on the test set for 100 repetitions are reported in
Table \ref{table: MALT}. In each of the 100 repetitions, we calculate the difference between the MSE of S-CMC$_\mathcal{H}$ and that of each competing method. Then we conducted a one-sided paired two-sample t-test with the alternative hypothesis that our method S-CMC$_\mathcal{H}$ has a smaller average mean squared error (MSE). Our method has the smallest MSE and
the $p$-values of the paired $t$-test indicates S-C$\mdc$
outperforms each method in prediction accuracy.
\begin{table}
\centering{}%
\begin{tabular}{lcc}
\hline 
Method & MSE & paired $t$-test $p$-value\tabularnewline
\hline 
CSIS & 1.364 & $<2.2\times10^{-16}$\tabularnewline
CDC-SIS & 1.162 & $<2.2\times10^{-16}$\tabularnewline
CIS& 1.268 & $<2.2\times10^{-16}$\tabularnewline
$\scmdc$ &0.748  & --\tabularnewline
\hline 
\end{tabular}\caption{The prediction accuracy in MALT data. \label{table: MALT} }
\end{table}
\subsubsection{Riboflavin Production Dataset\label{Rib-data}}

The dataset \citep{lee2001rna} contains information about riboflavin
(vitamin B2) production by $n=71$ bacillus subtiliswith, where $p$
= 4088 gene expression levels are recorded. The dataset is provided
by Royal DSM (Switzerland) and is available in the R package \textit{$\mathrm{\mathtt{hdi}}$}. 

Our goal is to find which genes are most related in predicting the riboflavin
production rate. We randomly split the sample into a training set
of size 60 and a test set of size 11. To evaluate the prediction performance
of each screening method, we select $d_2=16$ variables for each screening
method and train a random forest model on the training set. Then
we calculate the mean squared error (MSE) of each method on the test
set. The average MSE's based on 100 data splittings into training/test sets are reported in
Table \ref{table: Riboflavin}. Similar to Section \ref{realdata1}, we conducted the same one-sided paired two-sample t-test with the alternative hypothesis that our method S-CMC$_\mathcal{H}$ has a smaller averge mean squared error (MSE).Our method has the smallest MSE and
the $p$-values suggest that it significantly improves the MSE.

\begin{table}
\centering{}%
\begin{tabular}{lcc}
\hline 
Method & MSE & paired $t$-test $p$-value\tabularnewline
\hline 
MDC & 0.301 & $0.017$\tabularnewline
CSIS & 0.357 & $2.909\times10^{-15}$\tabularnewline
CDC-SIS & 0.323 & $5.62\times10^{-5}$\tabularnewline
CIS&0.323&$5.62\times10^{-5}$\tabularnewline
RaSE$_{1}$-eBIC & 0.293 & 0.401\tabularnewline
$\scmdc$ & 0.292 & --\tabularnewline
\hline 
\end{tabular}\caption{The prediction accuracy in Riboflavin production data. The conditional
set $\boldsymbol{X}_{S}$ contains the top $d_{1}=4$ variables suggested
by $\protect\mdc$.\label{table: Riboflavin} }
\end{table}

\section{Conclusion\label{section:conclusion}}

In this article, we propose the {\bf c}onditional {\bf m}artingale difference {\bf d}ivergence ($\cmdd$) to measure the dependence between
a response variable and a predictor vector given a third vector. It is primarily designed to overcome the limitation of marginal independence measures. Based on $\cmdd$, we develop
a new screening procedure called $\scmdc$ by combining the merits of the $\cmdc$ and $\mdc$ for selecting both marginal and jointly active variables. The proposed framework can be easily extended to
quantile screening. The simulations and real data applications demonstrate that $\scmdc$
has a competitive and stable performance under variety model settings for mean or quantile screening. We also provide a data-driven method for selecting the conditional set $\boldsymbol{X}_{S}$. The limitation of this method is that
we do need to predetermine a proper number of variables in conditional set to get a satisfactory performance. Using $\lfloor\sqrt{n/\log n}\rfloor$, as done in our numerical study, may
not suffice for the cases when the true underlying model consists jointly only active variables that depends on a large conditional set. Designing a variable screening method that is free of tuning the cardinality of the conditional
set is a challenging future research
topic.
\bigskip{}

\section*{Supplementary Materials}

The supplementary material
contains additional simulation results, and the proofs of theorems and results introduced in the main
article. (.pdf file)
\par
\section*{Acknowledgements}

The authors gratefully acknowledge the funding from \textit{National
Science Foundation (NSF-CIF-1813330)}.
\par




\bibhang=1.7pc
\bibsep=2pt
\fontsize{9}{14pt plus.8pt minus .6pt}\selectfont
\renewcommand\bibname{\large \bf References}
\expandafter\ifx\csname
natexlab\endcsname\relax\def\natexlab#1{#1}\fi
\expandafter\ifx\csname url\endcsname\relax
  \def\url#1{\texttt{#1}}\fi
\expandafter\ifx\csname urlprefix\endcsname\relax\def\urlprefix{URL}\fi

\bibliographystyle{apalike}
\bibliography{MDDH-Sinica}


\vskip .65cm
\noindent
Department of Statistics, University of Kentucky, Lexington KY 40536-0082, U.S.A. 
\vskip 2pt
\noindent
E-mail: (lfa246@uky.edu)
\vskip 2pt

\noindent
Department of Statistics, University of Kentucky, Lexington KY 40536-0082, U.S.A. 
\vskip 2pt
\noindent
E-mail: (chenglong.ye@uky.edu)

\bibhang=1.7pc
\bibsep=2pt
\fontsize{12}{14pt plus.8pt minus .6pt}\selectfont
\section{Appendix}

\appendix

\section{Bochner's Theorem}

\begin{lemma} \citep[Theorem 6.6]{wendland2004scattered}\label{thm1}
A continuous function $k:\mathbb{R}^{p+q}\rightarrow\mathbb{R}$ is
positive semi-definite if and only if it is the Fourier transform
of a finite nonnegative Borel measure $W(\xi)d\xi$ on $\mathbb{R}_{^{p+q}}$,
that is, 
\begin{align*}
k(z)=\int_{\mathbb{R}^{(p+q)}}e^{-iz^{T}\xi}W(\xi)d\xi,~\forall z\in\mathbb{R}^{p+q}.
\end{align*}
\end{lemma} 

\section{Properties of $\mdc$}
Some important
properties of $\mdd$ and $\mdc$ are
presented as follows.
\begin{definition} Given i.i.d. observations $(\boldsymbol{U}_{i},V_{i})_{i=1}^{n}$
from the distribution of $(\boldsymbol{U},V)$. Let $a_{ij}=V_{i}V_{j}$
and $b_{ij}=k(\boldsymbol{U}_{i}-\boldsymbol{U}_{j})$, where $k$ is a kernel in RKHS. The unbiased sample RKHS type martingale
difference divergence $\textnormal{\ensuremath{\mddn}}^2(V, \boldsymbol{U})$
is defined as 
\begin{align}
\textnormal{\ensuremath{\mddn}}^{2}(V, \boldsymbol{U})=\dfrac{1}{n(n-3)}\sum_{i\ne j}^{n}a_{ij}^{*}b_{ij}^{*}
\end{align}
 and the unbiased sample RKHS type martingale difference
correlation $\textnormal{\ensuremath{\mdcn}}^{2}(V,\boldsymbol{U})$ is defined by 
\begin{align}
\textnormal{\ensuremath{\mdcn}}^{2}(V,\boldsymbol{U})=\begin{cases}
\dfrac{\textnormal{\ensuremath{\mddn}}^{2}(V,\boldsymbol{U})}{\textnormal{var}_{n}(V)\textnormal{var}_{n\mathcal{H}}(\boldsymbol{U})} & \text{if \textnormal{var}\ensuremath{_{n}(V)}\textnormal{var}\ensuremath{_{n\mathcal{H}}(\boldsymbol{U})>0}}\\
0 & \text{ otherwise,}
\end{cases}
\end{align}
where var$_{n}(V)=(\frac{1}{n(n-3)}\sum_{i\ne j}^{n}|a_{ij}^{*}|^{2})^{1/2}$,
and var$_{n\mathcal{H}}(\boldsymbol{U})=(\frac{1}{n(n-3)}\sum_{i\ne j}^{n}|b_{ij}^{*}|^{2})^{1/2}$.
\end{definition}
\begin{theorem}\label{theorem:2} The following properties hold if $E(V^{2})<\infty$ : 
\begin{enumerate}
\item[a.] $\textnormal{MD}_{\mathcal{H}}^{2}(V,\boldsymbol{U})=E[(V-E(V))(V'-E(V'))k(\boldsymbol{U}-\boldsymbol{U}')]$. 
\item[b.]  $0\le\textnormal{MC}_{\mathcal{H}}(V,\boldsymbol{U})\le1,$
and $\textnormal{MC}_{\mathcal{H}}(V,\boldsymbol{U})=0\Leftrightarrow E(V|\boldsymbol{U})=E(V)$
almost surely. 
\item[c.] $\textnormal{MC}_{\mathcal{H}}(a+bV,c+\boldsymbol{U})=\textnormal{MC}_{\mathcal{H}}(V,\boldsymbol{U})$
for any scalars $a$, $b \in \mathbb{R}$ and $c\in\mathbb{R}^{q}$. 
\end{enumerate}
\end{theorem}
\begin{theorem} If E $(V^{2})<\infty$, then
\begin{equation}
lim_{n\rightarrow\infty}\textnormal{\ensuremath{\mddn}}(V,\boldsymbol{U})=\textnormal{\ensuremath{\mdd}}(V,\boldsymbol{U})\textrm{ a.s.,}
\end{equation}
and 
\begin{equation}
lim_{n\rightarrow\infty}\textnormal{\ensuremath{\mdcn}}(V,\boldsymbol{U})=\textnormal{\ensuremath{\mdc}}(V,\boldsymbol{U})\textrm{ a.s.}.
\end{equation}
\end{theorem}

\begin{theorem}

Assume $E(V^{2})<\infty$,
we have the following: 
\begin{enumerate}
\item[a.] If $\textnormal{MC}_{\mathcal{H}}(V,\boldsymbol{U})=0$, then 
\begin{equation}
n\textnormal{\ensuremath{\mddn}}^{2}(V,\boldsymbol{U})\xrightarrow[n\rightarrow\infty]{D}||\Gamma(s)||_{\mathcal{H}_k}^{2},
\end{equation}
where $\Gamma(\cdot)$ denotes a complex-valued zero-mean Gaussian
random process with covariance function 
\begin{align*}
\textnormal{cov}_{\Gamma}(s,s_{0})&=F(s-s_{0})-\textsl{g}_{\boldsymbol{U}}(s-s_{0})E^{2}(V)+\{E(V^{2})+E^{2}(V)\}\textsl{g}_{\boldsymbol{U}}(s)\overline{\textsl{g}_{\boldsymbol{U}}(s_{0})}\\
&-F(s)\overline{\textsl{g}_{\boldsymbol{U}}(s_{0})}-\textsl{g}_{\boldsymbol{U}}(s)\overline{F(s_{0})}
\end{align*}
 with $s,s_{0}\in\mathbb{R}^{q}$ and $g_{\boldsymbol{U}}(s)=E(e^{i\langle s,\boldsymbol{U}\rangle})$,
$F(s)=E[V^{2}\exp(i\langle \boldsymbol{U},s\rangle)]$. 
\item[b.] If $\textnormal{\ensuremath{\mdc}}(V,\boldsymbol{U})=0$ and $E(V^{2}|\boldsymbol{U})=E(V^{2})$,
then 
\[
n\textnormal{\ensuremath{\mddn}}^{2}(V,\boldsymbol{U})/S_{n}\xrightarrow[n\rightarrow\infty]{D}Q,
\]
where $S_{n}=(1-\frac{1}{n(n-1)}\sum_{k\ne l}k(\boldsymbol{U}_{k}-\boldsymbol{U}_{l}))(\frac{1}{n}\sum_{k}(V_{k}-\bar{V}_{n})^{2})$,
and $Q$ is a nonnegative quadratic form $Q=\sum_{i=1}^{\infty}\lambda_{i}Z_{i}^{2}$,
where $Z_{i}$ are independent standard normal random variables. $\{\lambda_{i}\}$
are nonnegative constants that depend on the distribution of (U,V)
and $E(Q)=1$. 
\item[c.] If $\textnormal{\ensuremath{\mdc}}(V,\boldsymbol{U})\ne 0$, then $n\textnormal{\ensuremath{\mddn}}^{2}(V,\boldsymbol{U})/S_{n}\xrightarrow[n\rightarrow\infty]{P}\infty$. 
\end{enumerate}
\end{theorem}

\subsection {Sure Screening Property of $\mdc$}\label{mdc}

Let $\psi_i=\mdc(Y,X_i)$ for predictor $X_i$ and $\widehat{\psi_i}=\mdcn(Y,X_i)$. Denote $\mathcal{\widehat{M
}}=\{j: \widehat{\psi}_j  \ge cn^{-\kappa},$ for $1\le j\le p \}$. Similar to $\cmdc$, we need the following two assumptions.

\begin{description}
\item [{(B1)}] There exists a positive constant $s_0$ such that for all $0<s\le 2s_0$, then
$ E\{$exp$(sY^2)\}<\infty$.


\item [{(B2)}] The minimum $\mdc$ value of active predictors is greather than $2cn^{-\kappa}$, for some constant $c>0$ and $0\le \kappa <\dfrac{1}{2}$.
\end{description}

\begin{theorem}
\label{theorem: mdc-SIS}
Under Assumption $(B1)$, for any $0<\gamma<1/2-\kappa$, there exist postive constants $c_1$ and $c_2$ such that 
\begin{equation}
P\Bigg\{\underset{1 \leq j \leq p}{\max} |\widehat{\psi}_j-\psi_j|\ge cn^{-\kappa}\Bigg\}\le O(p[\exp\{-c_1n^{1-2(\kappa+\gamma)}\}+n\exp(-c_2n^{\gamma})]). 
\end{equation}

Under conditions $(B1)$ and $(B2)$, we have that 
\begin{equation}
P(\mathcal{M}\subset \widehat{\mathcal{M}})\ge 1-O(s_n[\exp\{-c_1n^{1-2(\kappa+\gamma)}\}+n\exp(-c_2n^{\gamma})]),
\end{equation}
where $s_n$ is the cardinality of $\mathcal{M}$.

\end{theorem}

For the sure screening property of quantile screening by $\mdc$, we require the condition (C1) in the following Section \ref{sec:c}. Denote
$\mathcal{M}_{q_\tau}:=\{j:\mathbb{E}(Y_\tau|X_{j})\ \text{depends on}\ X_{j}\}$ and $\mathcal{\widehat{M
}}_{q_\tau}:=\{j: \widehat{\psi}_j(\hatw)  \ge cn^{-\kappa},$ for $1\le j\le p \}$.

\begin{theorem}
\label{theorem: mdc-SISQ}
	Under $(C1)$, for any $0<\gamma <1/2-\kappa$ and $\kappa \in (0,1/2)$, there exists positive constants $c_1, c_2$ such that for any $c>0$,
	\begin{equation}
	P\Bigg\{\underset{1 \leq j \leq p}{\max} |\widehat{\psi}_j(\hatw)-\psi_j(\myw)|\ge cn^{-\kappa}\Bigg\}\le O(p[\exp\{-c_1n^{1-2(\kappa+\gamma)}\}+n\exp(-c_2n^{\gamma})]).
	\end{equation}
	
If the minimum $\mdc$ value of active predictors satisfies
$\min_{j\in\mathcal{M}_{q_\tau}}\psi_{j}(\myw)\ge2cn^{-\kappa}$ for some
constant $c>0$ and $0\le\kappa<1/2$, we can show that 
\begin{equation}
P(\mathcal{M}_{q_\tau}\subseteq \widehat{\mathcal{M}}_{q_\tau})\ge 1-O(\widetilde s_n[\exp\{-c_1n^{1-2(\kappa+\gamma)}\}+n\exp(-c_2n^{\gamma})]),
\end{equation}
\end{theorem}

where $\widetilde{s_n}$ is the cardinality of $\mathcal{M}_{q_\tau}$.
\section{Sure Screening Property of Quantile Screening using $\cmdc$}\label{sec:c}

We require the two assumptions below:
\begin{description}
\item [{(C1)}] The CDF of $Y$ ($F_{Y}$) is continuously differentiable in
a small neighborhood of $q_{\tau}=q_{\tau}(Y)$, say $[q_{\tau}-\delta_{0},q_{\tau}+\delta_{0}]$
for $\delta>0$. Let $G_{1}(\delta_{0})=\inf{}_{y\in[q_{\tau}-\delta_{0},q_{\tau}+\delta_{0}]}f_{Y}(y)$,
and $G_{2}(\delta_{0})=\sup{}_{y\in[q_{\tau}-\delta_{0},q_{\tau}+\delta_{0}]}f_{Y}(y)$
where $f_{Y}$ is the density function of $Y$. Assume that $0<G_{1}(\delta_{0})\le G_{2}(\delta_{0})<\infty$.
\item [{(C2)}] The minimum C$\mdc$ value of active predictors satisfies
$\min_{j\in\mathcal{D}_{q_\tau}}\omega_{j}(\myw)\ge2cn^{-\kappa}$ for some
constant $c>0$ and $0\le\kappa<1/2$. 
\end{description}
The following proposition from \cite{shao2014martingale} is necessary for proving the sure screening
property. \begin{proposition} Under condition (C1),
there exists $\epsilon_{0}>0$ and $c_{1}>0$, such that for any $\epsilon\in(0,\epsilon_{0})$,
\begin{equation}
P\bigg(\dfrac{1}{n}\sum_{l=1}^{n}|\hat{y}_{l_\tau}-y_{l_\tau}|>\epsilon\bigg)\le3\exp(-2nc_{1}\epsilon^{2})
\end{equation}
\end{proposition}

\end{document}